\documentclass[conference]{IEEEtran}
\IEEEoverridecommandlockouts

\usepackage{graphicx}	
\usepackage{algorithm}
\usepackage{multirow}
\usepackage{booktabs}
\usepackage{paralist}
\usepackage{balance}
\usepackage[normalem]{ulem}
\useunder{\uline}{\ul}{}
\usepackage[noend]{algpseudocode}
\usepackage{cite,url}
\usepackage{amsmath,amssymb,amsfonts}
\usepackage{setspace}
\usepackage[justification=centering]{caption}
\usepackage{diagbox}
\usepackage{xcolor}
\usepackage{colortbl}
\usepackage{subcaption}
\usepackage[shortlabels]{enumitem}
\setenumerate[1]{itemsep=0pt,partopsep=0pt,parsep=\parskip,topsep=0.5pt}
\setitemize[1]{itemsep=0pt,partopsep=0pt,parsep=\parskip,topsep=0.5pt}
\setdescription{itemsep=0pt,partopsep=0pt,parsep=\parskip,topsep=0.5pt}

\let\temp\rmdefault
\usepackage{mathpazo}
\let\rmdefault\temp

\newcommand{\best}[1]{\textbf{#1}}
\newcommand{\secBest}[1]{\uline{#1}}
\newcommand\ExpCaption[1]{%
     \captionsetup{font=small}%
     \caption{#1}}
\newcommand\fptitle[1]{%
    \noindent\textbf{#1}{.}%
}     
\newcommand\ptitle[1]{%
    \noindent\textbf{#1}{.}%
}

\usepackage{amsthm}

\newtheoremstyle{exampstyle}
  {.2em} 
  {.2em} 
  {\itshape} 
  {} 
  {\bfseries} 
  {.} 
  {.5em} 
  {} 
\theoremstyle{exampstyle} \newtheorem{example}{Example}
\theoremstyle{exampstyle} 
\theoremstyle{exampstyle} 
\theoremstyle{exampstyle} 
\theoremstyle{exampstyle}
\newtheorem*{theorem*}{Problem}

\newcommand\AKM{\textsf{DasaKM}}
\newcommand\TAC{\textsf{TopoAC}}
\newcommand\ELKM{\textsf{ElbowKM}}

\newcommand\CD{\textsf{CD}}
\newcommand\LI{\textsf{LI}}
\newcommand\SL{\textsf{SL}}
\newcommand\MICE{\textsf{MICE}}
\newcommand\MF{\textsf{MF}}
\newcommand\BRITS{\textsf{BRITS}}
\newcommand\SSGAN{\textsf{SSGAN}}
\newcommand\BISIM{\textsf{BiSIM}}
\newcommand\NBISIM{\textsf{N-BiSIM}}
\newcommand\ABISIM{\textsf{D-BiSIM}}
\newcommand\TBISIM{\textsf{T-BiSIM}}
\newcommand\BISIMFamily{\textsf{*-BiSIM}}

\newcommand\SIV{Kaide}
\newcommand\SV{Wanda}
\newcommand\SVI{Longhu}

\newcommand\KNN{\textsf{$K$NN}}
\newcommand\WKNN{\textsf{W$K$NN}}
\newcommand\RF{\textsf{RF}}

\newcommand\MNAR{MNAR}
\newcommand\MAR{MAR}
\newcommand\NULL{\texttt{null}}

\newcommand\DA{DA}
\newcommand\APE{APE}

\newcommand{\Harry}[1]{{\textcolor{blue} {#1}}}
\newcommand{\Xiao}[1]{{\textcolor{red} {#1}}}

\newcommand{\change}[1]{{\textcolor{blue} {#1}}}

\usepackage[colorinlistoftodos,prependcaption,textsize=tiny]{todonotes}
\newcommand{\rone}[2][1=]{\todo[linecolor=blue,backgroundcolor=blue!25,bordercolor=blue]{#2}}
\newcommand{\rthree}[2][1=]{\todo[linecolor=orange,backgroundcolor=orange!25,bordercolor=orange]{#2}}
\newcommand{\rfour}[2][1=]{\todo[linecolor=olive,backgroundcolor=olive!25,bordercolor=olive]{#2}}

\def\BibTeX{{\rm B\kern-.05em{\sc i\kern-.025em b}\kern-.08em
    T\kern-.1667em\lower.7ex\hbox{E}\kern-.125emX}}
\begin{document}

\title{Data Imputation for Sparse Radio Maps in Indoor Positioning (Extended Version)}


\author{
Xiao Li$^1$ \hspace{1em} Huan Li$^2$ \hspace{1em} Harry Kai-Ho Chan$^3$ \hspace{1em} Hua Lu$^1$ \hspace{1em} Christian S. Jensen$^4$\\
$^1$Department of People and Technology, Roskilde University, Denmark \\
$^2$College of Computer Science and Technology, Zhejiang University, China \\
$^3$Information School, University of Sheffield, United Kingdom \\
$^4$Department of Computer Science, Aalborg University, Denmark \\
 $^1$\{xiaol, luhua\}@ruc.dk \hspace{1em} $^2$lihuan.cs@zju.edu.cn \hspace{1em} $^3$h.k.chan@sheffield.ac.uk \hspace{1em} $^4$csj@cs.aau.dk
}

\maketitle


\begin{abstract}
Indoor location-based services rely on the availability of sufficiently accurate positioning in indoor spaces. A popular approach to positioning relies on so-called radio maps that contain pairs of a vector of Wi-Fi signal strength indicator values (RSSIs), called a fingerprint, and a location label, called a reference point (RP), in which the fingerprint was observed. The positioning accuracy depends on the quality of the radio maps and their fingerprints. Radio maps are often sparse, with many pairs containing vectors missing many RSSIs as well as RPs. Aiming to improve positioning accuracy, we present a complete set of techniques to impute such missing values in radio maps. 
We differentiate two types of missing RSSIs: missing not at random (MNAR) and missing at random (MAR). Specifically, we design a framework encompassing a missing RSSI differentiator followed by a data imputer for missing values. The differentiator identifies MARs and MNARs via clustering-based fingerprint analysis. 
Missing RSSIs and RPs are then imputed jointly by means of a novel encoder-decoder architecture that leverages temporal dependencies in data collection as well as correlations among fingerprints and RPs. A time-lag mechanism is used to consider the aging of data, and a sparsity-friendly attention mechanism is used to focus attention score calculation on observed data. Extensive experiments with real data from two buildings show that our proposal outperforms the alternatives with significant advantages in terms of imputation accuracy and indoor positioning accuracy.

\end{abstract}

\pagestyle{plain}

\section{Introduction}
\label{sec:intro}
Indoor applications involving navigation, augmented reality, and moving robots require sufficiently accurate indoor positioning. 
According to Research and Markets, the global indoor positioning and navigation market will exceed \$54 billion by 2026~\cite{indoorPositioning}.
While a variety of indoor positioning technologies exist, positioning based on Wi-Fi fingerprinting~\cite{he2015wi} is popular: the ubiquity of Wi-Fi enables positioning without the deployment of additional expensive infrastructure, and the technology is non-intrusive to users.
However, the accuracy of Wi-Fi fingerprinting based positioning depends heavily on the quality of the radio map data used~\cite{quezada2022data,lin2020locater, xu2015enhancing,sun2021data}.

\begin{figure}[!ht]
\centering
\includegraphics[width=\columnwidth]{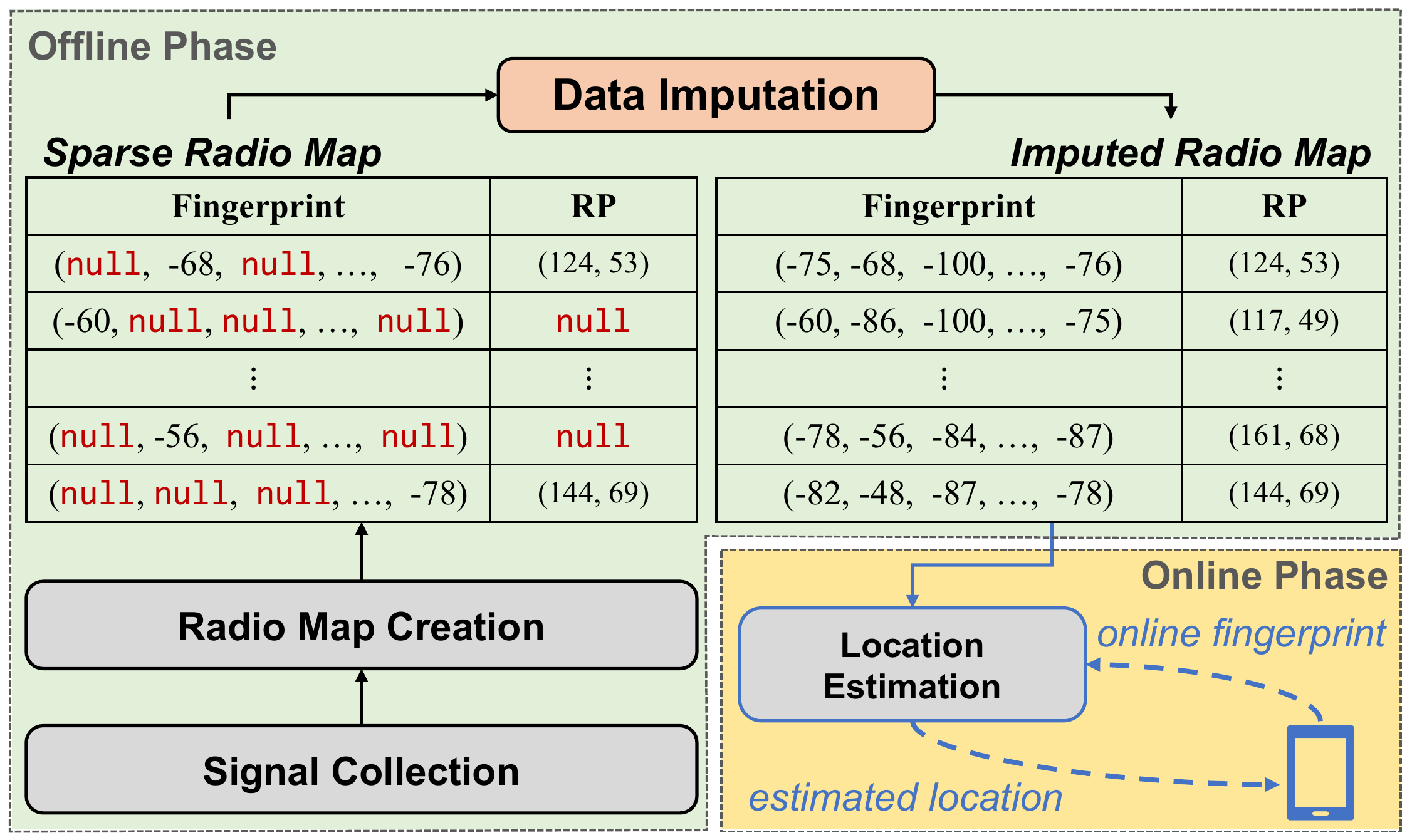}
\caption{Fingerprinting procedure and radio maps.}
\label{fig:missing_rate_raw}
\end{figure}

Wi-Fi fingerprinting entails two phases, as shown in Fig.~\ref{fig:missing_rate_raw}. The offline phase creates a so-called radio map that contains pairs of a vector of Wi-Fi \textit{received signal strength indicator} values (RSSIs), called a fingerprint, 
and a location label,
called a reference point (RP), in which the fingerprint was observed. An RSSI measures the signal strength of a Wi-Fi \textit{access point} (AP)~\cite{jung2016performance}, and is an integer value in the range of $[-99,0]$ dBm. An example radio map is shown in the top-left part of Fig.~\ref{fig:missing_rate_raw}. The online phase localizes the users by utilizing a location estimation algorithm (e.g., \KNN{}~\cite{zeinalipour2017anatomy}) that compares the user device's fingerprint with the radio map.

\vspace{5pt}
\vspace{5pt}

\begin{figure}[!ht]
\centering
\includegraphics[width=\columnwidth]{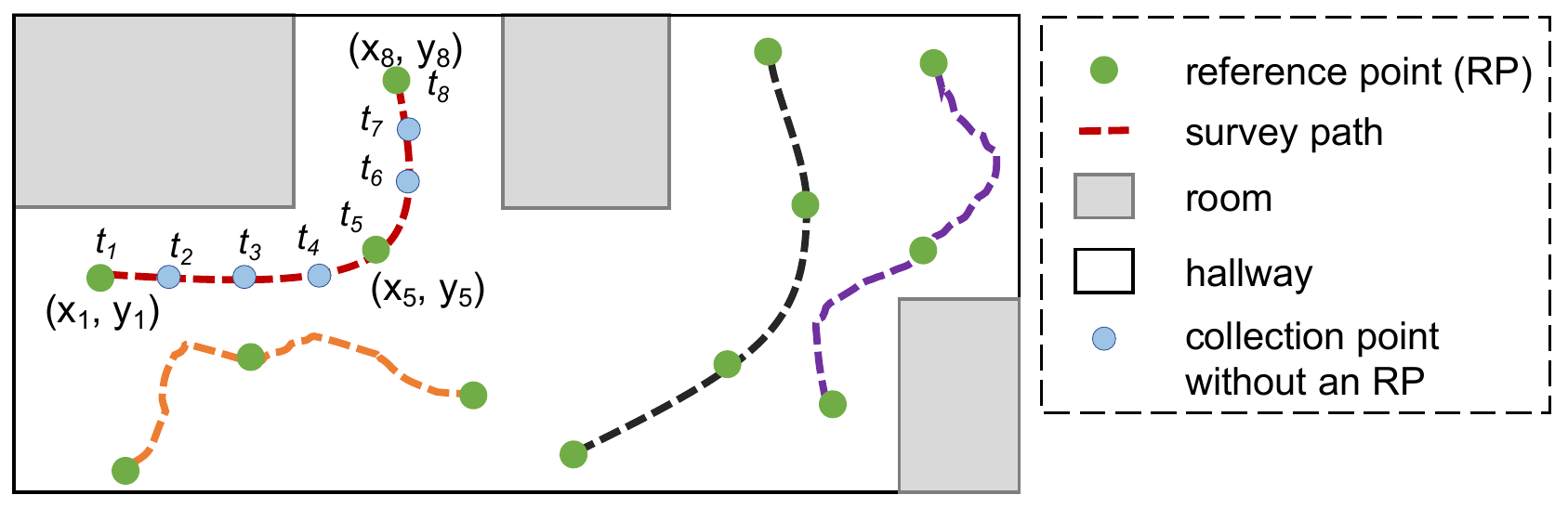}
\caption{Walking survey based data collection.}
\label{fig:floorplan}
\end{figure}

To build radio maps efficiently and economically for an indoor space, surveys are often performed by surveyors moving in the indoor space~\cite{jung2016performance,han2014kailos,li2017turf,chang2014crowdsourcing, wu2014smartphones}. 
{Surveyors collect RSSIs continuously while moving along predefined paths, as illustrated in Fig.~\ref{fig:floorplan}.}
%
Due to fluctuation in the wireless environment and asynchrony between 
{the collection} 
and RPs (to be detailed in Section~\ref{ssec:radio_map}), 
\vspace{5pt}
\vspace{5pt}
the results of walking surveys
 suffer from low data quality,
having
high rates of missing  (i.e., percentages of \NULL{}s)
%
%
 RSSIs and RPs in a radio map.
For example, in the radio maps obtained from walking surveys
 in  two real buildings called \SIV{} and \SV{} (to be detailed in Section~\ref{ssec:settings}),
the  rates of missing RSSIs and RPs are between 85.6\% and 93.7\%.
%
In other words, the radio maps are highly sparse, having many \NULL{}s. 
%
Such \NULL{}s must be replaced by real numbers in order for the radio map to be used by location estimation algorithms~\cite{kaiser2014dealing, sorour2014joint}.
Intuitively, imputing accurate real numbers for \NULL{}s in a radio map  improves its usability in indoor positioning. However, existing studies  employ  straightforward strategies to fill-in RP \NULL{}s~\cite{firdaus2019accurate,dong2017dealing,gorak2018automatic} and RSSI \NULL{}s~\cite{kaiser2014dealing,li2017turf,sorour2014joint}, yielding subpar results (cf.\ Section~\ref{ssec:evaluation_data_imputation}). Therefore, this study focuses on improving the quality and usability of sparse radio maps by accurately imputing missing RSSI and RP data.
In doing so, the study contends with difficult challenges.

First, two types of missing RSSIs exist.
\textbf{Missing Not At Random} (\MNAR{}) RSSIs are caused by the unobservability of the signals of APs. This typically occurs when an AP is too far away and cannot be seen by a user's device. 
In contrast, \textbf{Missing At Random} (\MAR{}) RSSIs\footnote{The terms MNAR and MAR stem from literature~\cite{rubin1976inference, little2019statistical}. They are applied to missing RSSI values in this work.} result from random events, e.g., the temporary presence of obstacles in transmission paths or occasional loss of contact with APs~\cite{firdaus2019accurate,dong2017dealing}. 

\begin{figure}[!ht]
    \centering
    \begin{minipage}[t]{0.236\textwidth}
    \centering 
    \includegraphics[width=\textwidth]{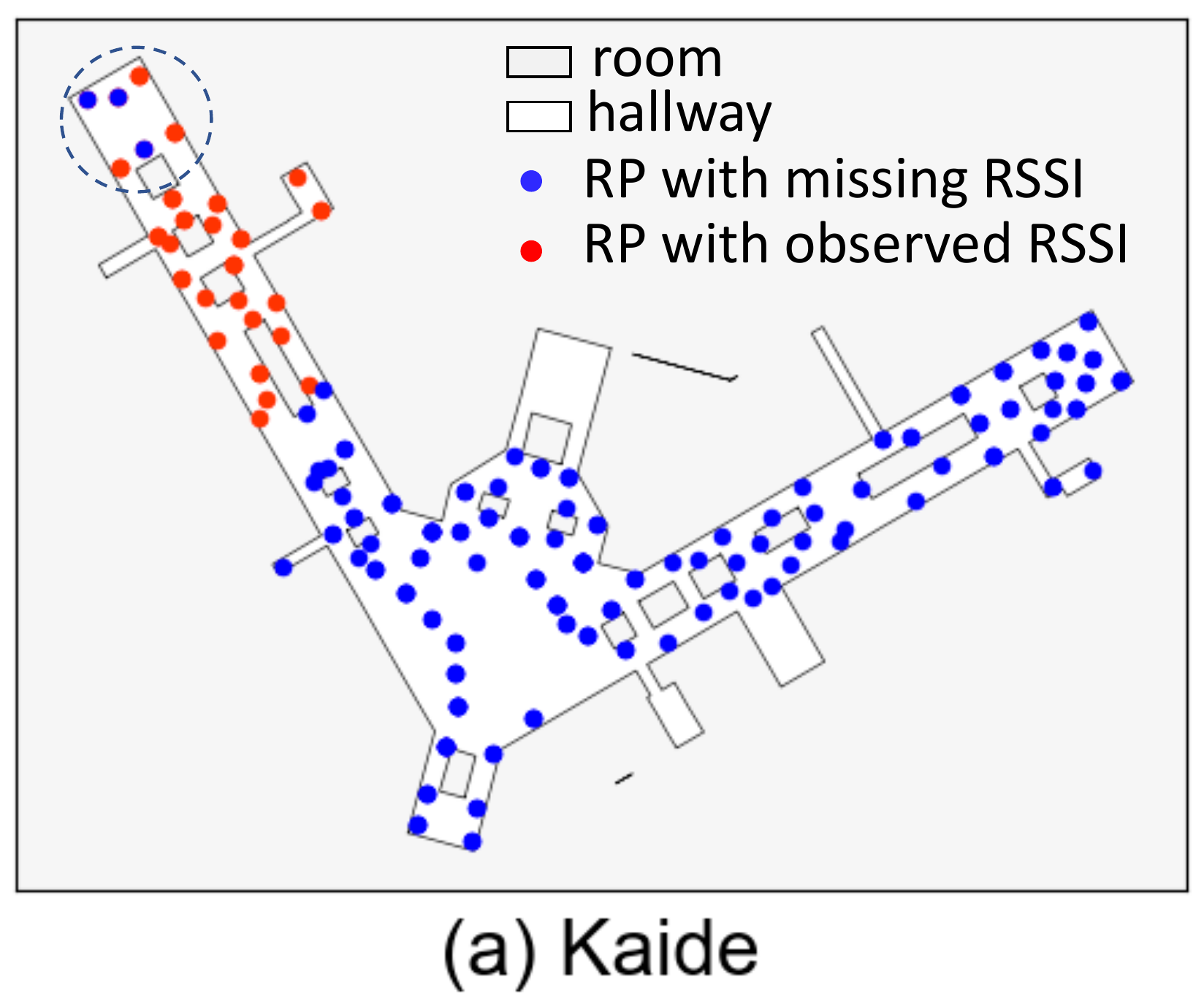}
    \end{minipage}
    \begin{minipage}[t]{0.236\textwidth}
    \centering
    \includegraphics[width=\textwidth]{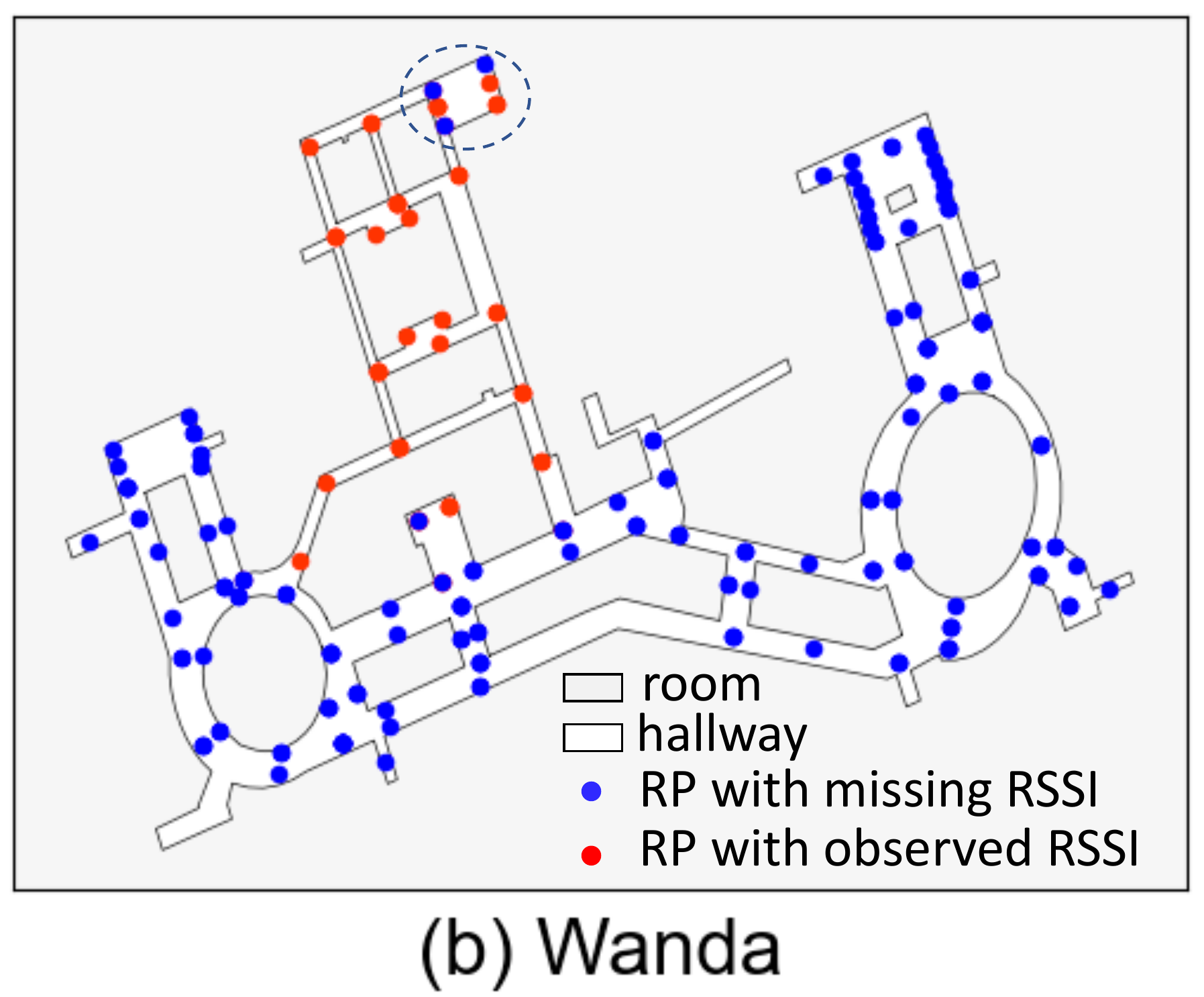}
    \end{minipage}
    \caption{Observability of a selected AP's signals at different reference points (RPs).}\label{fig:mar_mnar}
\end{figure}

An example of \MAR{} RSSI and \MNAR{} RSSI is shown in Fig.~\ref{fig:mar_mnar}. An AP (access point) is selected in each venue and its deployment location is roughly within the dashed circle.
For an RP (reference point), if all fingerprints collected at that RP have observed the selected AP, the RP is marked in red; otherwise, some of its fingerprints have missed the selected AP, and that RP is marked in blue.
Clearly, most RPs far away from the selected AP are blue, indicating that the selected AP is unobservable at these RPs and the corresponding missing events are classified as Missing Not At Random (MNAR).
On the other hand, most of the RPs near the dashed circle are red but there are several blue RPs that sometimes miss the selected AP's signals. The missing events in these RPs are incidental and should be treated as Missing At Random (MAR).
\vspace{5pt}
\vspace{5pt}

%
%
%
As the two types of missing RSSIs have different causes and meanings, they should be differentiated before imputation~\cite{rubin1976inference}. 
However, neither traditional radio map completion methods~\cite{gorak2018automatic,li2017turf,pulkkinen2011semi} nor general data imputers~\cite{hastie2009elements,azur2011multiple, che2018recurrent, yoon2017multi,cao2018brits, miao2021generative, cini2021multivariate,bansal2021missing} differentiate the missing RSSI types. The former simply assume that all missing RSSIs are \MNAR{}s, while the latter treat them as \MAR{}s. We will offer empirical  evidence of the benefits of differentiation.

To differentiate MNARs and MARs, we design a clustering based differentiator that clusters a radio map’s fingerprints, and determines \MAR{}s and \MNAR{}s via  intra-cluster analyses. To obtain appropriate clusters, we design two clustering algorithms that utilize a specialized accuracy metric and indoor topology, respectively.

The subsequent imputation of missing values is also challenging.
Following existing radio map completion studies~\cite{gorak2018automatic,li2017turf,pulkkinen2011semi}, we replace identified \MNAR{} values by the value $-100$ dBm, the lowest RSSI value that thus reflects the unobservability of \MNAR{}s.
However, imputing \MAR{}s and missing RPs is not straightforward.
A \MAR{} should be imputed with a value
in $[-99, 0]$ dBm,\footnote{$-99$ dBm $\gg$ $-100$ dBm in terms of power as dBm is log-based~\cite{dBm}.} since its value would have been observed had the random event that caused it not occurred.
Traditional radio map completion methods impute missing RPs using linear interpolation~\cite{li2017turf} or semi-supervised learning~\cite{sorour2014joint}.
General data imputation methods employ matrix 
factorization~\cite{hastie2009elements} or chained equations~\cite{azur2011multiple}.
However, all these methods fall short when data sparsity is high, and they do not consider the correlations and temporal dependencies between  RSSIs and RPs collected along a path in walking-survey based radio map data collection.
%
%
Next, time-series imputation methods~\cite{che2018recurrent, yoon2017multi,cao2018brits, miao2021generative, cini2021multivariate,bansal2021missing} target missing values in feature or source sequences (e.g., multivariate time series) with known labels. In contrast, we must contend with  heterogeneous missing values---\MAR{}s in source (or fingerprint) sequences and missing RPs in target (or RP) sequences. 

To impute missing \MAR{}s and RPs effectively for sparse radio maps, we design an encoder-decoder based data imputer that exploits both temporal dependencies in time series  and correlations between source and target sequences to impute \MAR{}s in source sequences and missing RPs in target sequences jointly. {A standard encoder-decoder is unsuitable in this setting, due to the many missing values in both source (fingerprint)  and  target (RP) sequences. The irregularity in the absence of RSSIs results in different durations between consecutive encoder units (see Table~\ref{tab:input_feature} in Section~\ref{sec:imputation}-B), which a standard encoder-decoder cannot handle. Further, missing values in the input also degrades the ability of the model's attention mechanism at capturing the importance of each unit.
To tackle these issues, we  introduce a time-lag mechanism that considers the aging of the last observed value when modeling the relationships between  consecutive encoder units, and we design an adapted attention mechanism to contend with the high sparsity of input features.}

We make the following major contributions.
\begin{itemize}[leftmargin=*]
    \item We differentiate missing RSSIs as \MAR{}s and \MNAR{}s. To the best of our knowledge, we are the first to do so. Specifically, we provide a clustering-based approach to differentiate  missing RSSIs (Section~\ref{sec:differentiation}). 
    
    

    \item We devise a novel encoder-decoder that is capable of imputing missing RSSIs and RPs jointly by exploiting temporal dependencies in fingerprint and RP sequences as well as  correlations among fingerprints and RPs. The data imputer considers both the aging of missing values and the sparsity of input sequences (Section~\ref{sec:imputation}).

    \item We report on extensive experiments on real data, finding that our proposals outperform the alternatives substantially in terms of data imputation accuracy and indoor positioning accuracy (Section~\ref{sec:exp}). 
\end{itemize}

Section~\ref{sec:prel} presents  preliminaries and problem settings, Section~\ref{sec:related} reviews  related work, and Section~\ref{sec:conclusion} concludes and discusses future work.


\section{Preliminaries and Problem Settings}
\label{sec:prel}

Table~\ref{tab:notation} lists  notations used in the paper.

\begin{table}[!htbp]
\centering
\small
{\setlength\tabcolsep{6pt}
\caption{Notation}
\label{tab:notation}
\begin{tabular}{c|l}
\toprule
{Symbol} & {Description} \\ \midrule
$r_d$ & RSSI of the $d$th AP \\
$\mathbf{f} = ( r_1, r_2, \ldots, r_D )$ & a  fingerprint  of RSSIs from $D$ APs \\
$\mathbf{l} = ( x, y ) $ & a location or a reference point (RP) \\
$\{ (\mathbf{f}_i, \mathbf{l}_i): i = 1 \text{~to~} N \}$ & a radio map $ \in \mathbb{R}^{N \times (D+2)}$ \\
$\mathbf{M} \in \{-1,0,1\}^{N \times D}$ & a radio map mask matrix \\
$\mathbf{b}_i$ & a binary RSSI profile vector of $\mathbf{f}_i$ \\
$\mathbf{x}_i = \mathbf{b}_i \oplus \mathbf{l_i}$ & a concatenated radio map sample \\
\bottomrule
\end{tabular}}
\end{table}

\subsection{Fingerprinting based Indoor Positioning}
\label{ssec:fingerprinting}

Given $D$ Access Points (APs), a Wi-Fi \textbf{fingerprint} $\mathbf{f} = ( r_1, r_2, \ldots, r_D )$ is a vector of one
\textbf{received signal strength indicator value (RSSI)} per AP as measured at
 a  \textbf{reference point (RP)}, so that $r_d$ is the RSSI of the $d$th AP. The location $\mathbf{l} = (x, y)$ of an RP is usually preselected by a surveyor.
A radio map consists of $N$ pairs of the form, i.e.,  $(\mathbf{f}_i,\mathbf{l}_i)$, where $\mathbf{f}_i$ is the fingerprint obtained at location $\mathbf{l}_i$.

For simplicity, we consider a single floor. 
In a multi-floor setting, our proposal can be applied to each floor separately, as studies show that it is possible to perform floor identification  with high accuracy (e.g.,  99+\%~\cite{wang2017research}).

As mentioned, fingerprinting based positioning has two phases. 
In the offline phase,  surveyors collect fingerprints  and use the collected data to create a radio map. 
%
We target the relatively efficient data collection approach based on walking surveys, to be detailed in Section~\ref{ssec:radio_map}.

In the online {location estimation} phase,  a user's current location is estimated by  an algorithm that compares an online fingerprint $\mathbf{f}_o$ from the user's device with a pre-collected radio map.
Typical location estimation algorithms are listed below. 
\begin{itemize}[leftmargin=*]
\item \KNN{}~\cite{zeinalipour2017anatomy} finds $\mathbf{f}_o$'s $K$ nearest fingerprints in the radio map and uses the mean of their RPs as the estimated location.
\item Unlike \KNN{}, \WKNN{}~\cite{fang2008location} uses a weighted mean.  Weights are inversely proportional to the distances between $\mathbf{f}_o$ and the fingerprints in the radio map.
\item 
Others~\cite{jedari2015wi} use a radio map (fingerprints as features and RPs as labels) to train a regression model
(e.g., a Random Forest) that predicts $\mathbf{f}_o$'s location.
\end{itemize}
In all cases, the positioning accuracy relies heavily on the radio map data quality.


\subsection{Walking Survey based Radio Map Creation}
\label{ssec:radio_map}

In a walking survey~\cite{jung2016performance,han2014kailos,li2017turf,chang2014crowdsourcing, wu2014smartphones}, a surveyor visits a sequence of preselected RPs with flexible movement in-between each two consecutive RPs, collecting RSSIs of APs along with corresponding collection times and then enters these into a \emph{Walking Survey Record Table}.

Fig.~\ref{fig:floorplan} shows an example with four survey paths.
 The top-left one yields the record table  in Table~\ref{tab:wifi_raw_data}.
There are two types of records, namely RP and RSSI, sorted on timestamps. 
The surveyor started at RP $(x_1, y_1)$ at time $t_1$, visited RP $(x_5, y_5)$ at $t_5$, and reached RP $(x_8, y_8)$ at $t_8$.
The RSSI records capture additional RSSI data, e.g., at time $t_2$, RSSIs of the 1st, 2nd, and 3rd APs are $-70$ dBm, $-83$ dBm and $-76$ dBm, respectively.

As it is possible that the two types of records are collected asynchronously, a pre-processing method has been widely used~\cite{sun2020wifi} to create the radio map as follows.
\begin{itemize}[leftmargin=*]
\item \textbf{Step 1} merges  consecutive RSSI records if their time difference is below a threshold $\epsilon$.
The merged record uses the earlier time, and gets its RSSIs as follows. 
If an AP is in one record only, that RSSI is used. If an AP is in both records, the average RSSI is used. Otherwise, \NULL{} is used.
\item \textbf{Step 2} merges  consecutive RSSI and RP records if their times differ by less than $\epsilon$. The time and RSSIs are as produced in Step 1; the RP is copied from the RP record. Each remaining RSSI or RP record is converted into a record in which each missing value is set to \NULL{}.

\end{itemize}

The threshold $\epsilon$ is specified by the surveyor.
Setting $\epsilon$ = 1 for Table~\ref{tab:wifi_raw_data}, we get the radio map records and  times  in Table~\ref{tab:incmp_radio_map}.
Though a radio map does not contain timestamps, we show them in Table~\ref{tab:incmp_radio_map} because we use them for imputation later on.
In Step 1, RSSI records at $t_6$ and $t_7$ are merged into $\langle r_1:-74, r_2:-77, r_3:\NULL{}, r_4: \NULL{}, r_5:-81 \rangle$ at $t_6=12$. 
In Step 2, this new record is not merged with an RP record but is converted to a pair $(( -74, -77, \NULL{}, $ $\NULL{}, -81),$ $\NULL{})$.
In contrast, the RP record at $t_1$ is merged with the RSSI record at $t_2$, resulting in the pair $(( -70, -83, -76, \NULL{}, \NULL{} ),$ $( x_1, y_1 ))$.
Likewise, records at $t_4$ and $t_5$ are merged into $(( \NULL{}, \NULL{},$ $ -80, -68, \NULL{} ), ( x_5,y_5 ))$.
Moreover, the RP record at $t_8$ is converted to $(( \NULL{}, \NULL{},$ $ \NULL{}, \NULL{},$ $\NULL{} ),$ $( x_8, y_8 ))$.

\begin{table}[t]
    \centering
    {\setlength\tabcolsep{1.2pt} 
    \caption{Walking Survey Record Table}
    \label{tab:wifi_raw_data}
    \footnotesize
    \begin{tabular}{@{}ccc|ccc@{}}
    \toprule         Time & Type & Measurement & Time & Type & Measurement \\
\midrule
         $t_1=0$ & RP & $(x_1, y_1)$  & $t_5=9$ & RP & $(x_5, y_5)$\\
         $t_2=1$ & RSSI & $\langle r_1:-70, r_2:-83, r_3:-76 \rangle$ & $t_6=12$ & RSSI &$\langle r_1:-74, r_5:-80\rangle$\\
         $t_3=3$ & RSSI & $\langle r_1:-71, r_3:-78 \rangle$  & $t_7=13$ & RSSI & $\langle r_2:-77, r_5:-82 \rangle$\\
         $t_4=8$ & RSSI & $\langle r_3:-80, r_4:-68\rangle$ & $t_8=16$ & RP & $(x_8, y_8)$\\
\bottomrule
    \end{tabular}}
\end{table}
\begin{table}[t]
    \centering
    \caption{Created Radio Map}
    \label{tab:incmp_radio_map}
    \small
    \begin{tabular}{c|c|c}
    \toprule
         No. & Radio Map Record & Time \\
         \midrule
         1 &  $(( -70, -83, -76, \NULL{}, \NULL{} ), ( x_1, y_1 ))$ & $t_2$\\
         2 & $(( -71, \NULL{}, -78, \NULL{}, \NULL{} ), \NULL{})$ & $t_3$\\
         3 & $(( \NULL{}, \NULL{}, -80, -68, \NULL{} ), ( x_5, y_5 ))$ & $t_4$ \\
         4 & $(( -74, -77, \NULL{}, \NULL{}, -81 ), \NULL{} )$ & $t_6$\\
         5 & $(( \NULL{},  \NULL{}, \NULL{}, \NULL{}, \NULL{} ), ( x_8, y_8 ))$ & $t_8$\\
        \bottomrule
    \end{tabular}
\end{table}

This method generates temporally dense radio map records that, however, may contain many RP and RSSI \NULL{}s. 

\subsection{Problem and Solution Overview}
\label{ssec:problem_solution}

\begin{theorem*}[Radio Map Imputation]
Given a radio map, we impute the  RSSI and RP   \NULL{} values in the radio map, such that indoor positioning using this radio map yields lower positioning errors.
\end{theorem*}

\begin{figure}
\centering
 \includegraphics[width=\columnwidth]{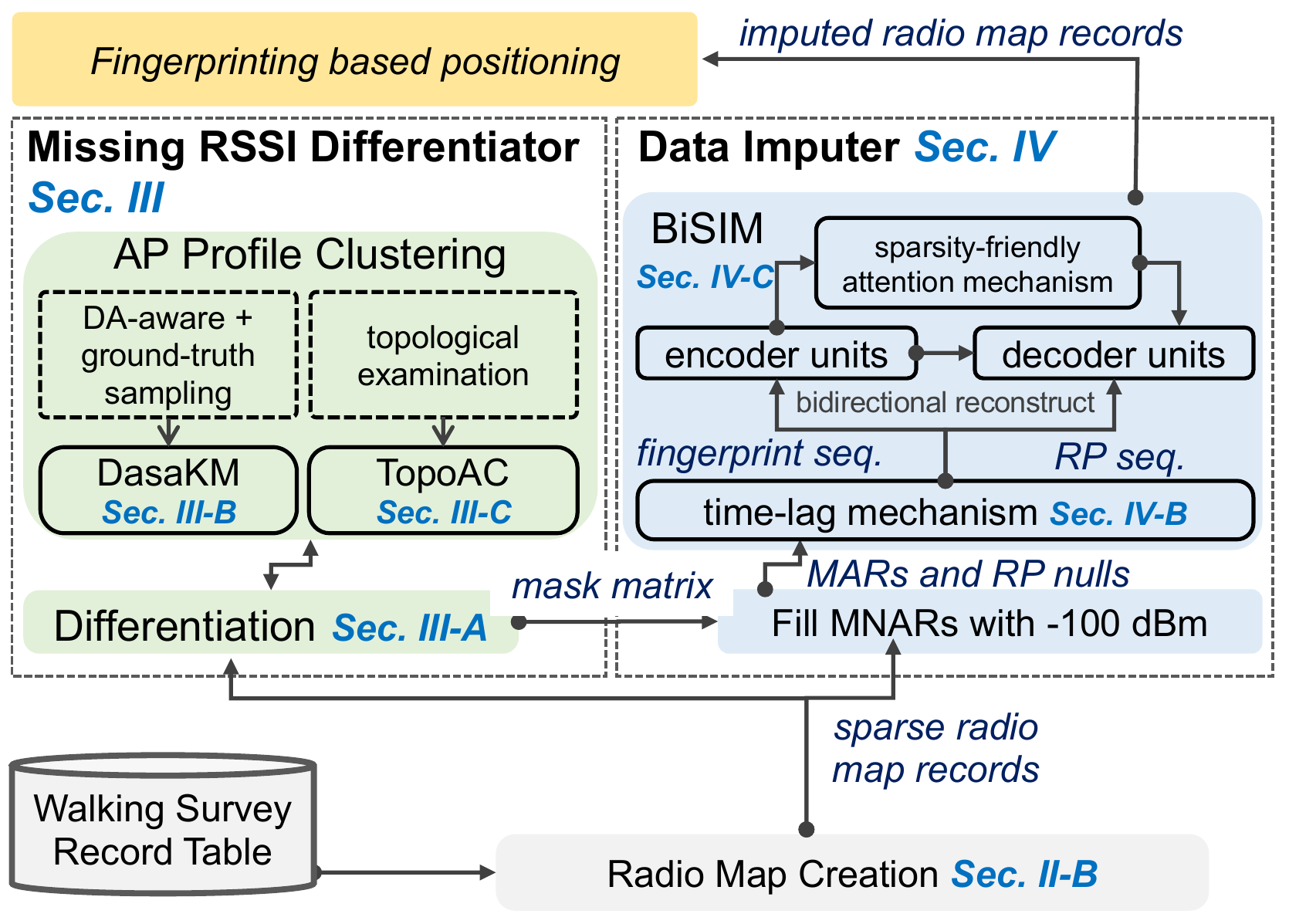}
\caption{Framework overview.}
\label{fig:framework}
\end{figure}

%
As pointed out in Section~\ref{sec:intro}, missing RSSIs are  random or non-random, yielding 
%
%
\emph{Missing At Random} (\MAR{}) and  \emph{Missing Not At Random} (\MNAR{}) RSSIs.
They should be differentiated before data imputation~\cite{rubin1976inference}. 
%
\if
In this study, we propose a radio map imputation framework that 1) differentiates \MNAR{}s and \MAR{}s for proper subsequent data imputation and 2) imputes correlated missing RPs and RSSIs simultaneously. The framework is illustrated in Figure~\ref{fig:framework}.
\Harry{The two modules in the framework are described briefly below.}
\fi
To solve this problem, we propose a framework (cf.\ Fig.~\ref{fig:framework}) with two modules.

\textbf{Missing RSSI Differentiator Module} (Section~\ref{sec:differentiation}). Given a radio map, this module categorizes missing RSSIs as  \MNAR{}s and \MAR{}s.
Specifically, the differentiation process (Section~\ref{ssec:differentiator}) regards \MAR{}s as random absences of AP signals in  fingerprints and employs a clustering based approach to identify those random absences according to the locality of AP profiles (i.e., observability of APs). We design two algorithms (Sections~\ref{ssec:akm} and~\ref{ssec:thac}) for clustering AP profiles in different ways. 
%
The differentiation process returns the recognized \MNAR{}s and \MAR{}s as a mask matrix, where $-1$ means \MNAR{}, $0$ means \MAR{}, and $1$ means 
an observed RSSI.

\textbf{Data Imputer Module} (Section~\ref{sec:imputation}) imputes  missing RSSI and RP values.
Initially, all  \MNAR{}s are assigned the value $-100$ dBm.
Then, a bidirectional encoder-decoder based model called \BISIM{} (Section~\ref{ssec:architecture}) imputes \MAR{} and RP \NULL{}s jointly for a sequence of radio map records from a survey path.
In particular, \BISIM{} considers the aging of records by applying a \emph{time-lag mechanism} (Section~\ref{ssec:bisim_input}) to sequential radio map records. \BISIM{}  subsequently encodes fingerprint feature sequences and decodes the corresponding RP feature sequences to capture  correlations in a radio map record and among sequential radio map records.
\BISIM{} also employs a \emph{sparsity-friendly attention mechanism} (Section~\ref{ssec:bisim_internals}) to perform weight calculation against missing values.
%
The fingerprints and RPs predicted sequentially by the encoder/decoder units form the final imputed radio map records.

\section{Missing RSSI Differentiator}
\label{sec:differentiation}
\subsection{Differentiation Approach}
\label{ssec:differentiator}

In a wireless setting, identifying \MNAR{}s is non-trivial due to the complexity of analyzing the signal transmission paths between RPs and APs~\cite{sadowski2018rssi}.
To this end, we instead identify \MAR{}s as ``unusual'' RSSI missing events when comparing to  observed RSSIs in the same or similar signal environments. 
We thus rely on the following \textbf{hypothesis}:
%
\textit{Within a certain small range of space, the observability of APs is similar due to the similar signal transmission surroundings.}

To verify this hypothesis, we did an exploratory analysis on two real-world shopping malls named \SIV{} and \SV{} that we describe in detail in Section~\ref{ssec:settings}.

First, we generate an \textbf{AP profile} for each observed RP
by a process called \textsc{Binarization}. The process of \textsc{Binarization} is shown in Algorithm~\ref{alg:binarization}. 
%
%
We assume each RP corresponds to one fingerprint. In case  multiple fingerprints are generated for an RP, the fingerprints are averaged into one.
The process constructs a $D$-dimensional binary vector $\mathbf{b}_i$ for the RP $\mathbf{l}_i$: 
$\mathbf{b}_i[d]=1$ if the $d$th AP is observed at $\mathbf{l}_i$, and $\mathbf{b}_i[d]=0$, otherwise.

\vspace{10pt}
\begin{algorithm}[!htbp]
\caption{\textsc{Binarization} (an RP $\mathbf{l_i}$'s fingerprint $\mathbf{f}_i$)}\label{alg:binarization}
\begin{algorithmic}[1]
\State binary vector $\mathbf{b}_i \gets \mathbf{1}^{D}$
\For {$d = 1$ to $D$}
    \If {$\mathbf{f}_i[d]$ is \NULL{}}
        $\mathbf{b}_i[d] \gets 0$
    \EndIf
\EndFor
\State \Return $\mathbf{b}_i$
\end{algorithmic}
\end{algorithm}
\vspace{5pt}

\begin{figure}[!ht]
    \centering
    \begin{minipage}[t]{0.236\textwidth}
    \centering 
    \includegraphics[width=\textwidth]{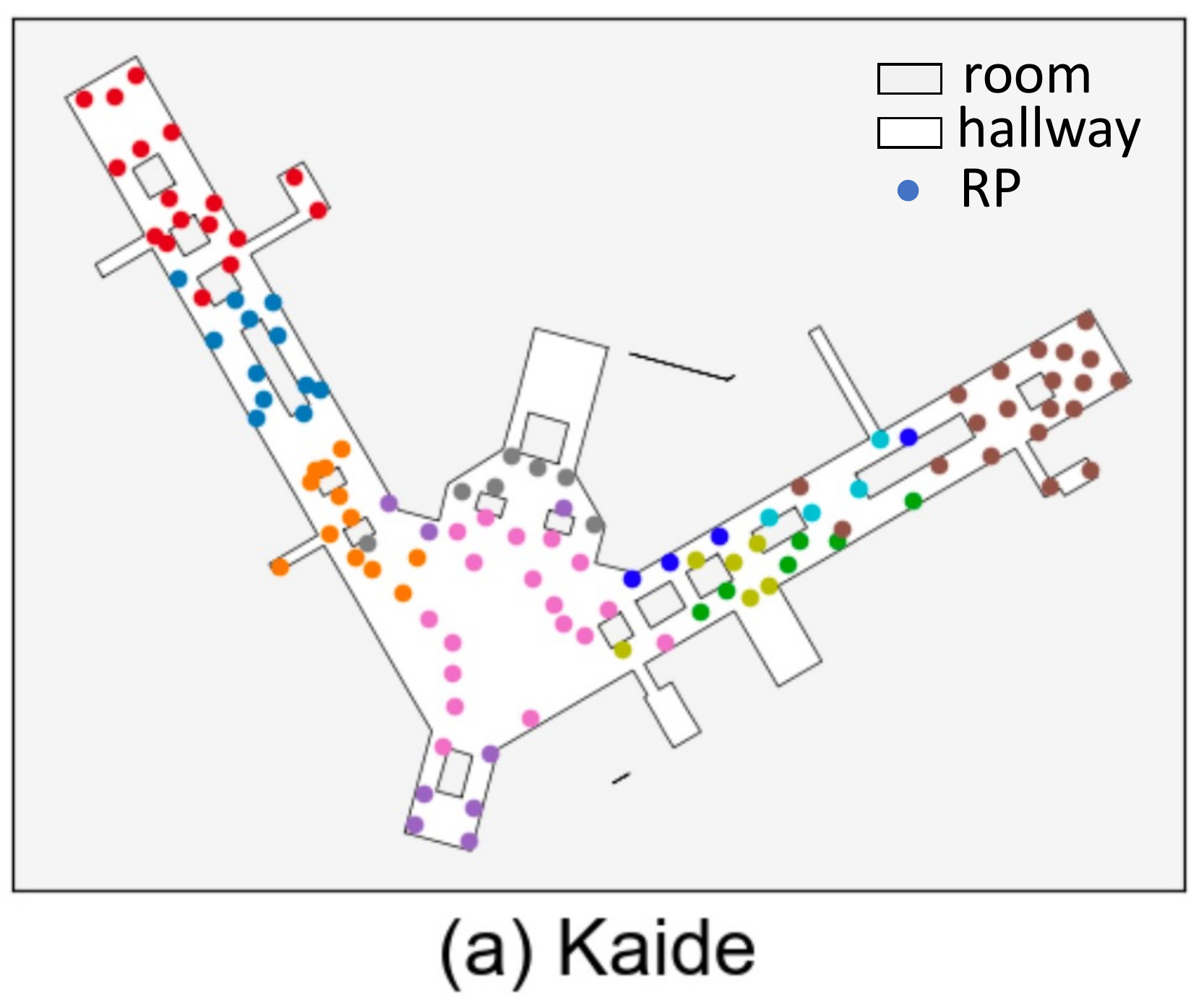}
    \end{minipage}
    \begin{minipage}[t]{0.236\textwidth}
    \centering
    \includegraphics[width=\textwidth]{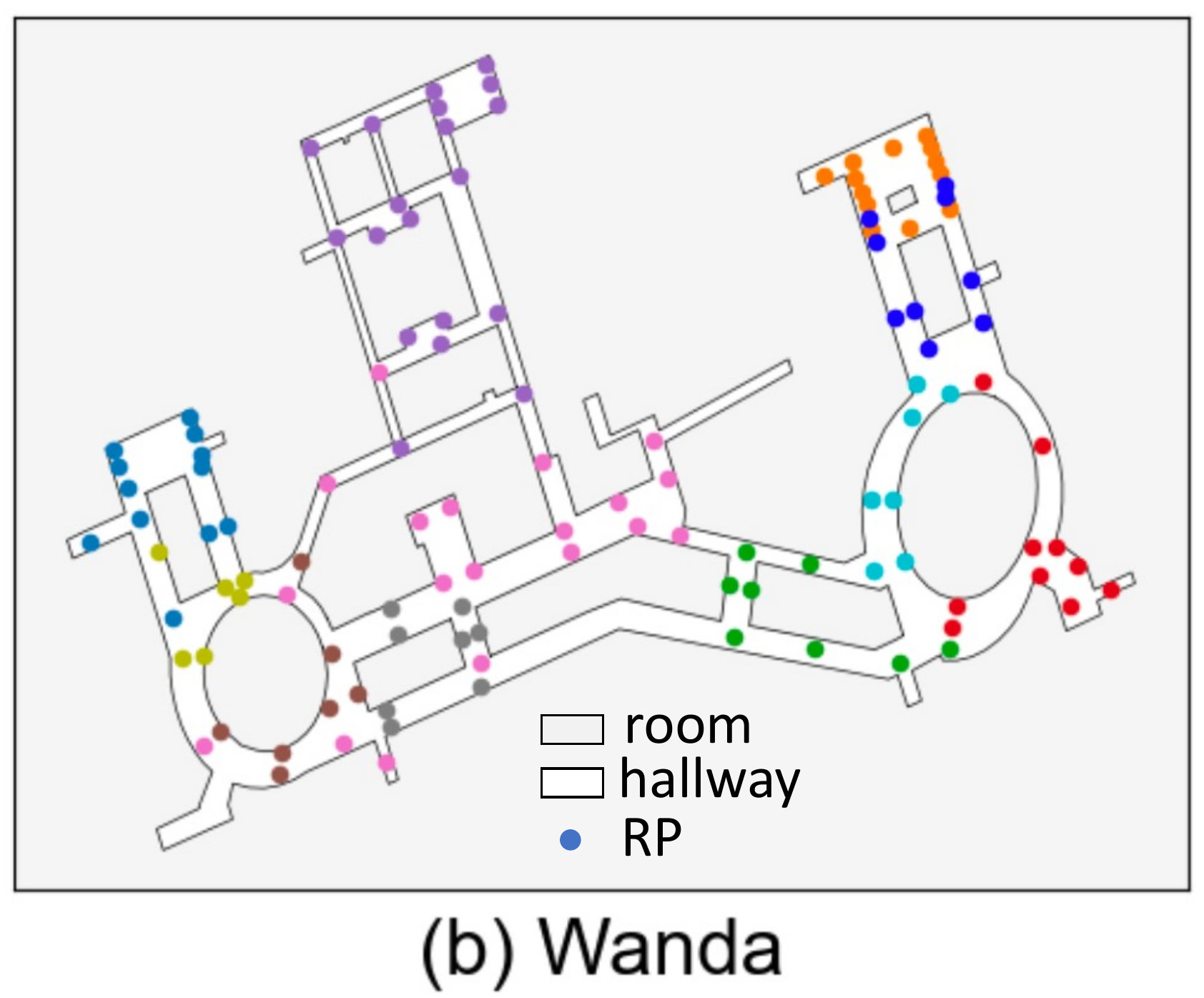}
    \end{minipage}
    \caption{Preliminary clustering tests on real-world venues.}\label{fig:clustering_test}
\end{figure}

Next, we conducted a clustering of the binarized AP profiles.
We use the widely-used $K$-means using Euclidean distance\footnote{We also considered Manhattan distance, but it achieved inferior results. We thus employ Euclidean distance for $K$-means unless  stated otherwise.} and tune the hyperparameter $K$ carefully.
We color the resulting clusters and visualize the RPs in Fig.~\ref{fig:clustering_test}.
We see that in most cases, the similar AP profiles (in the same cluster) are spatially close to each other.
Although some exceptions occur due to noise (\MAR{}s) in fingerprints when generating the AP profiles,
the hypothesis holds.

\vspace{5pt}
\vspace{5pt}

We thus identify \MAR{}s based on the clustering of AP profiles. The idea is that if a value $r_d$ is missing in an AP profile $\mathbf{b}_i$ while an $r_d$ value is present in a certain fraction of AP profiles similar to $\mathbf{b}_i$ in the same cluster, the missing $r_d$ is likely to be a \MAR{} in the fingerprint. To this end, a threshold $\eta$ is used such that a fraction higher than $\eta$ indicates \MAR{}s.

Algorithm~\ref{alg:differentiation} formalizes the differentiator with a predefined fraction threshold $\eta$ as the input. It returns an $N \times D$ mask matrix $\mathbf{M}$ (initialized in line~1), where $\mathbf{M}[i,j]$ is $0$ if the $j$th ($1 \leq j \leq D$) AP dimension of the $i$th fingerprint ($1 \leq i \leq N$) in the radio map
is a \MAR{}, $-1$ if it is an \MNAR{}, and $1$ if it is observed.
%
Lines~2--5 construct the sample set $X$ for clustering.
We highlight two differences related to $X$ in Algorithm~\ref{alg:differentiation} versus the exploratory analysis:
First, each sample in $X$ is a concatenation of the AP profile and the RP location.
This enables us to utilize  prior knowledge of RP locations to form clusters with spatially close RPs.
Second, $X$ covers all radio map records including those with \NULL{} RPs. To this end, each \NULL{} RP is  interpolated linearly based on its previously and subsequently observed RPs in the radio map.
Although imprecise, these interpolated RP positions capture  spatial proximity, which improves the clustering effectiveness. 
Line~6 generates a set $C$ of clusters by one of two clustering algorithms (to be detailed in Sections~\ref{ssec:akm} and~\ref{ssec:thac}).

Lines~7--12 identify \MAR{}s in each cluster $c_k$ by considering all its AP profiles.
In particular, each AP dimension $r_j$ is checked (line~8) to determine whether the missing of $r_j$ in $c_k$ is unusual. If $\eta_j$, the fraction of observed $r_j$ across all samples in $c_k$, exceeds the threshold $\eta$, $r_j$ \NULL{}s are \MAR{}s and marked as $0$ in $\mathbf{M}$. Otherwise, they are \MNAR{}s and marked as $-1$ (lines~9--12).
Finally, $\mathbf{M}$  is returned.

\vspace{10pt}
\begin{algorithm}
\caption{\textsc{Differentiation} (fraction threshold $\eta$)} \label{alg:differentiation}
\begin{algorithmic}[1]
\State mask matrix $\mathbf{M} \gets \mathbf{1}^{N \times D}$
\State sample set $X \gets \varnothing$
\For {each record $( \mathbf{f}_i, \mathbf{\hat{l}}_i )$ in the radio map}
    \State $\mathbf{x}_i \gets \textsc{Binarization}(\mathbf{f}_i) \oplus \mathbf{\hat{l}}_i$ \Comment{$\mathbf{\hat{l}}_i$ is  interpolated linearly}
    \State add $\mathbf{x}_i$ to $X$
\EndFor
\State $C \gets \textsc{Clustering}(X)$
\For {each cluster $c_k \in C$}
    \For {each AP dimension $r_j$}
        \State $\eta_j \gets$ the fraction of observed $r_j$ for all samples in $c_k$
        \If {$\eta_j > \eta$}
            \State mark all $r_j$ \NULL{}s within $c_k$ as $0$ in $\mathbf{M}$  \Comment{\MAR{}s}
        \EndIf
        \State \textbf{else} mark all $r_j$ \NULL{}s within $c_k$ as $-1$ in $\mathbf{M}$ \Comment{\MNAR{}s}
    \EndFor
\EndFor
\State \Return $\mathbf{M}$
\end{algorithmic}
\end{algorithm}
\vspace{10pt}

Algorithm~\ref{alg:differentiation} works with different clustering algorithms.
Section~\ref{ssec:akm} presents \AKM{} (Differentiation accuracy aware, sampling-based $K$-means) to replace the manually-tuned $K$-means used in the exploratory analysis.
%
In Section~\ref{ssec:thac}, we utilize indoor topology information and devise \TAC{} (Topology-aware  Agglomerative Clustering) that achieves even better performance without hyperparameters.
In Section~\ref{ssec:evaluation_differentiation}, 
Algorithm~\ref{alg:differentiation} is  evaluated experimentally with different clustering algorithms in terms of  indoor positioning error. 
In general, \TAC{} performs better as it takes the indoor topology into account, while \AKM{} does not require any prior knowledge.

\subsection{Algorithm \AKM{} }
\label{ssec:akm}

A straightforward way of applying $K$-means is to use the elbow method~\cite{kargar2021predict} that employs a \textit{within-cluster sum of square} metric to examine intra-cluster similarity. This method, however, leads to subpar performance at missing RSSI differentiation (see evaluations in Section~\ref{ssec:evaluation_differentiation}) as it disregards our ultimate goal of differentiation. To address this, we propose a more intuitive metric called differentiation accuracy (\DA{}) that measures the differentiation ability of the clustering result. As the ground-truth \MAR{}s and \MNAR{}s are not known, we first propose a ground-truth sampling procedure. 

\ptitle{Ground-truth Sampling Procedure}
It is non-trivial to generate the ground-truth mask matrix $\mathbf{M}_g$ by manually differentiating \MAR{}s and \MNAR{}s.
Thus, we modify the original sample set $X$ to ``create''  ground-truth \MAR{}s and \MNAR{}s:
\begin{itemize}[leftmargin=*]

\item \emph{Sampling \MAR{}s}. We nullify some observations in a record and mark them as $0$ in $\mathbf{M}_g$. They correspond to random RSSI missing events that are actually observable.

\item \emph{Sampling \MNAR{}s}. We search the indoor venue to sample a set of  adjacent RPs  that cover a sufficiently large area in the venue\footnote{In our implementation, we fix the RP size  to $6$. It forms a sufficiently large area and also avoids extra search cost caused by a larger size.}. These RPs are likely to share a similar AP profile.
If such RPs all missed an AP dimension in their records, then their corresponding missing values should be \MNAR{}s. The relevant masks in $\mathbf{M}_g$ are set to $-1$ accordingly.
\end{itemize}


Ideally, \MAR{}s and \MNAR{}s should be sampled according to their real distributions in the original dataset, which are, however, unknown.
Hence, to mitigate potential ground-truth sampling biases, we propose to sample multiple ground-truth sets using different proportions of \MAR{}s and \MNAR{}s and measure the average accuracy on these ground-truth sets.
Moreover, we design the differentiation accuracy as a balanced metric that is agnostic to the imbalanced proportion of the sampled ground-truth set.

\noindent\textbf{Differentiation Accuracy Metric}. 
The design of \DA{} is based on the metric called \emph{balanced accuracy}, which is shown to be effective for imbalanced positive and negative samples~\cite{altman1994diagnostic,garcia2010theoretical}. Specifically, \DA{} computes the true positive rate as the fraction of positive samples (\MAR{}s) identified correctly, and the true negative rate as the fraction of negative samples (\MNAR{}s) identified correctly.
Then, \DA{} simply takes the arithmetic average of the true positive rate and the true negative rate, thus disregarding the ratio of positive and negative ground-truth samples.
The arithmetic average used by \DA{} implies that either class is of equal importance for differentiation.
In contrast, the conventional  $F$-score measures only the performance of identifying positive samples---its \emph{precision} and \emph{recall} measure the fractions of correct positive samples in the result and positive samples being returned, respectively. Thus, the $F$-score is not used to implement \DA{}.

\noindent\textbf{Algorithm}. \AKM{} (Algorithm~\ref{alg:akm}) first  generates iteratively a ground-truth set ${GS}_{\gamma}$ in a particular input proportion of sampled \MAR{}s and \MNAR{}s and then removes it from the input dataset to form $X_{\gamma}$ (lines~1--3).
Next, it goes through a set of $K$ values until reaching a predefined upper-bound $U$ and selects the optimal $\hat{K}$ as the one achieving the highest \DA{} (lines~4--10). For each $K$, \DA{} is averaged over different ground-truth datasets (lines 6--9).
Finally, the $K$-means clustering on the original data $X$ using $\hat{K}$ is returned (line~11).
\vspace{10pt}
\begin{algorithm}
\caption{\textsc{\AKM} (sample set $X$, proportion  list $\Gamma$, upper-bound $U$)} \label{alg:akm}
\begin{algorithmic}[1]
\For {proportion $\gamma \in \Gamma$}
\State sample a ground-truth set ${GS}_{\gamma}$ from $X$ such that $\gamma = \frac{\text{\#(\MNAR{}s)}}{\text{\#(\MAR{}s)}}$ 
\State $X_{\gamma} \gets X \setminus$ ${GS}_{\gamma}$
\EndFor
\State $\mathit{maxDA} \gets 0$; $\hat{K} \gets 0$
\For {$K = 1$ to $U$} 
    \For{$\gamma \in \Gamma$} \Comment{try different sampled datasets}
    \State $C_{\gamma} \gets$ \textsc{KMeans}($X_{\gamma}$, $K$) 
    \State $\mathit{DA}_{\gamma} \gets$ calculate \DA{} w.r.t $C_{\gamma}$ and ${GS}_{\gamma}$
    \EndFor
    \State  $\widehat{DA} \gets \operatorname{average}( \{ \mathit{DA}_{\gamma} \mid \gamma \in \Gamma \})$
    \If {$\widehat{DA} > \mathit{maxDA}$}
        $\hat{K} \gets K$
    \EndIf
\EndFor
\State \Return \textsc{KMeans}($X$, $\hat{K}$)
\end{algorithmic}
\end{algorithm}
\vspace{10pt}

\begin{figure}[!ht]
\centering
    \begin{minipage}[t]{0.236\textwidth}
    \centering 
    \includegraphics[width=\textwidth]{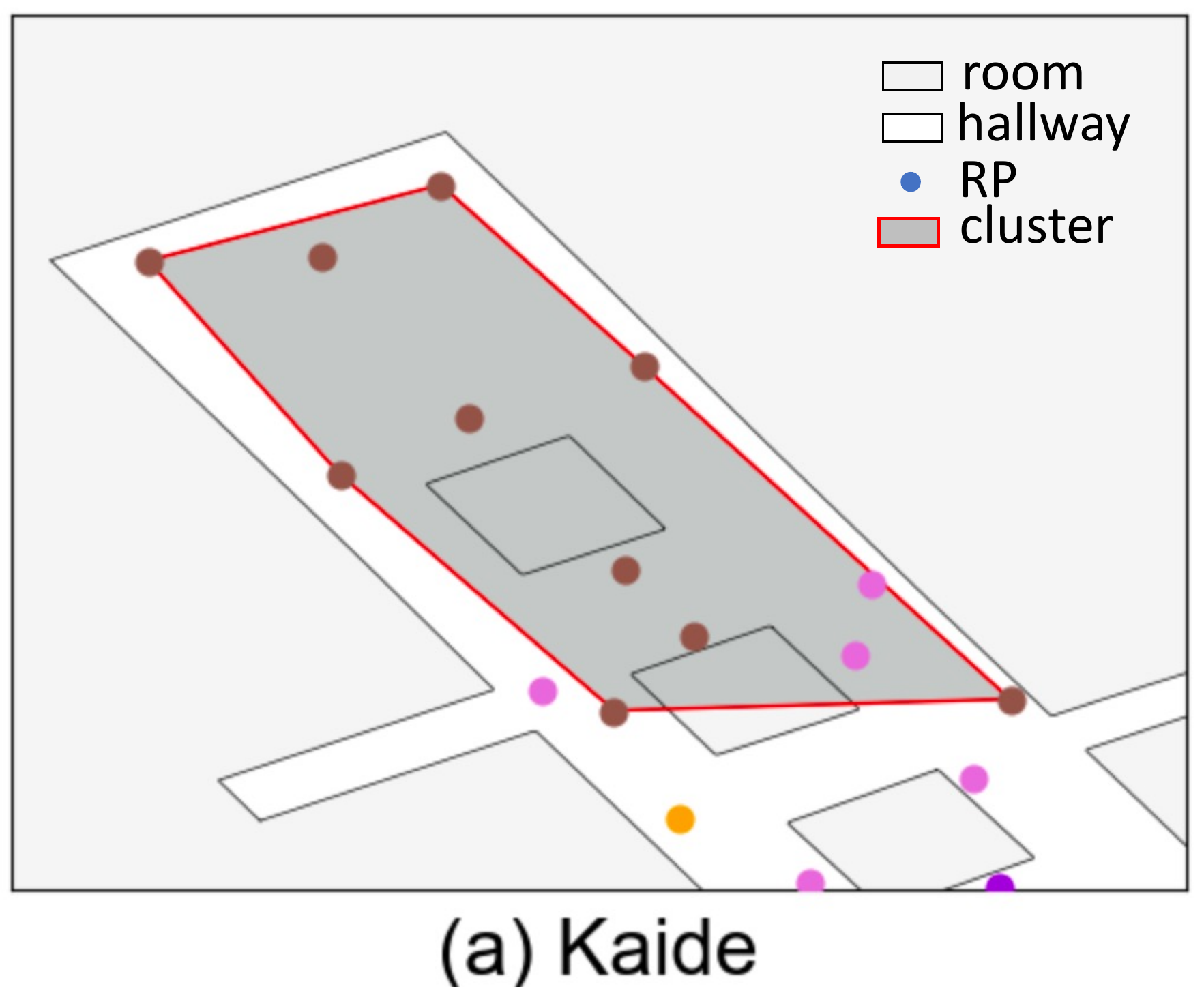}
    \end{minipage}
    \begin{minipage}[t]{0.236\textwidth}
    \centering
    \includegraphics[width=\textwidth]{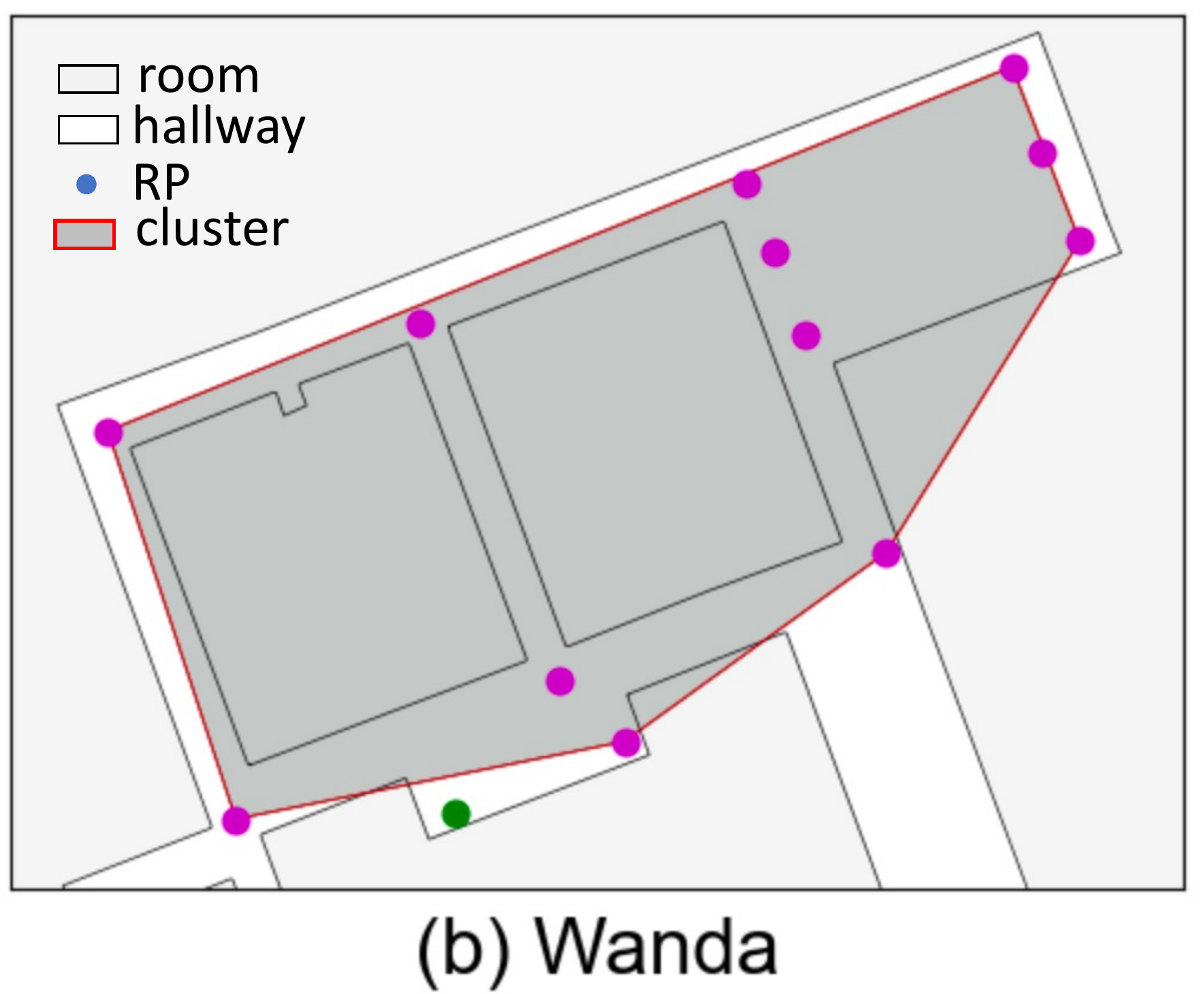}
    \end{minipage}
    \caption{Result of \AKM{}.}\label{fig:akm}
\end{figure}

%
\AKM{} finds close samples based on {inter-vector distances in a transformed signal space}, which may, however, batch  samples having distinct signal transmission surroundings in the indoor space.
We have found  two abnormal cases, shown as the two resultant clusters in Fig.~\ref{fig:akm}.
%
Their RPs scatter around the rooms, and their AP profiles may differ largely due to the existence of the walls among them which constitute distinct signal transmission environments.
%
\vspace{5pt}
\vspace{5pt}

\subsection{Algorithm \TAC{} }
\label{ssec:thac}

To avoid   abnormal cases and improve  accuracy, we design the Topology-aware Agglomerative Clustering (\TAC{}) that considers the  topology of the indoor space.

\fptitle{Heuristic of Topology}
If a set of RPs share similar AP profiles, there should not exist topological entities such as walls and obstacles that cause non-line-of-sight signal propagation within the closed region of these RPs.
In other words, if the convex hull of a set of RPs contains topological entities, these RPs should not form a cluster.
The basis for such a heuristic is formalized in Algorithm~\ref{alg:topology}. It takes as input a cluster $c$ and topological entities $\mathcal{T}$ in the form of a multipolygon.
It returns True if any entities exist in the convex hull $\mathit{CH}$ formed by the locations in  cluster $c$. Otherwise, it returns False. For example,  the case  in Fig.~\ref{fig:akm}(a)  returns True because that cluster $\mathit{CH}$  intersects  polygons in $\mathcal{T}$.

\vspace{10pt}
\begin{algorithm}
\caption{\textsc{EntityExist} (cluster $c$, multipolygon $\mathcal{T}$)} \label{alg:topology}
\begin{algorithmic}[1]
\State location set $L \gets \{l_i\ \mid \ x_i=(f_i,l_i) \wedge x_i\in c\}$
\State{$\mathit{CH} \gets$  convex hull covering $L$}
\State\Return $(\mathit{CH} \setminus \mathcal{T} \not= \varnothing)$
\end{algorithmic}
\end{algorithm}
\vspace{10pt}

\ptitle{Integrating the Topology Heuristic into the Algorithm}
The above topology heuristic can  be integrated naturally into an agglomerative clustering process where two adjacent clusters are merged  if the resulting cluster passes the examination of Algorithm~\ref{alg:topology}.
Note that the heuristic does not work with $K$-means, where it is too complex to assign samples to clusters while satisfying the heuristic.

The integrated clustering is detailed as 
\TAC{} in Algorithm~\ref{alg:thac}.
Initially, each sample $\mathbf{x}_i$ forms a single cluster $c_i$.
It then iteratively merges the pair of clusters with the minimum center-to-center Euclidean distance that passes the topological examination (lines 2--4).
It terminates when no  clusters can be merged. \TAC{} does not require any hyperparameters.

\vspace{10pt}
\begin{algorithm}
\caption{\textsc{\TAC{}} (sample set $X$, multipolygon $\mathcal{T}$)} \label{alg:thac}
\begin{algorithmic}[1]
\State initialize $C \gets \{ c_i \mid \text{for~each~} \mathbf{x}_i \in X \}$

\While{{$\exists$ cluster pair $(c_i, c_j)$ \textit{s.t.} $!\textsc{EntityExist}(c_i \cup c_j, \mathcal{T})$}}
     \State pick $(c_i', c_j')$ with the minimum distance  \textit{s.t.} $!\textsc{EntityExist}(c_i' \cup c_j', \mathcal{T})$
     \State merge $c_i'$ and $c_j'$ in $C$
\EndWhile
\State \Return $C$
\end{algorithmic}
\end{algorithm}
\vspace{10pt}

\begin{figure}[!ht]
\centering
    \begin{minipage}[t]{0.236\textwidth}
    \centering 
    \includegraphics[width=\textwidth]{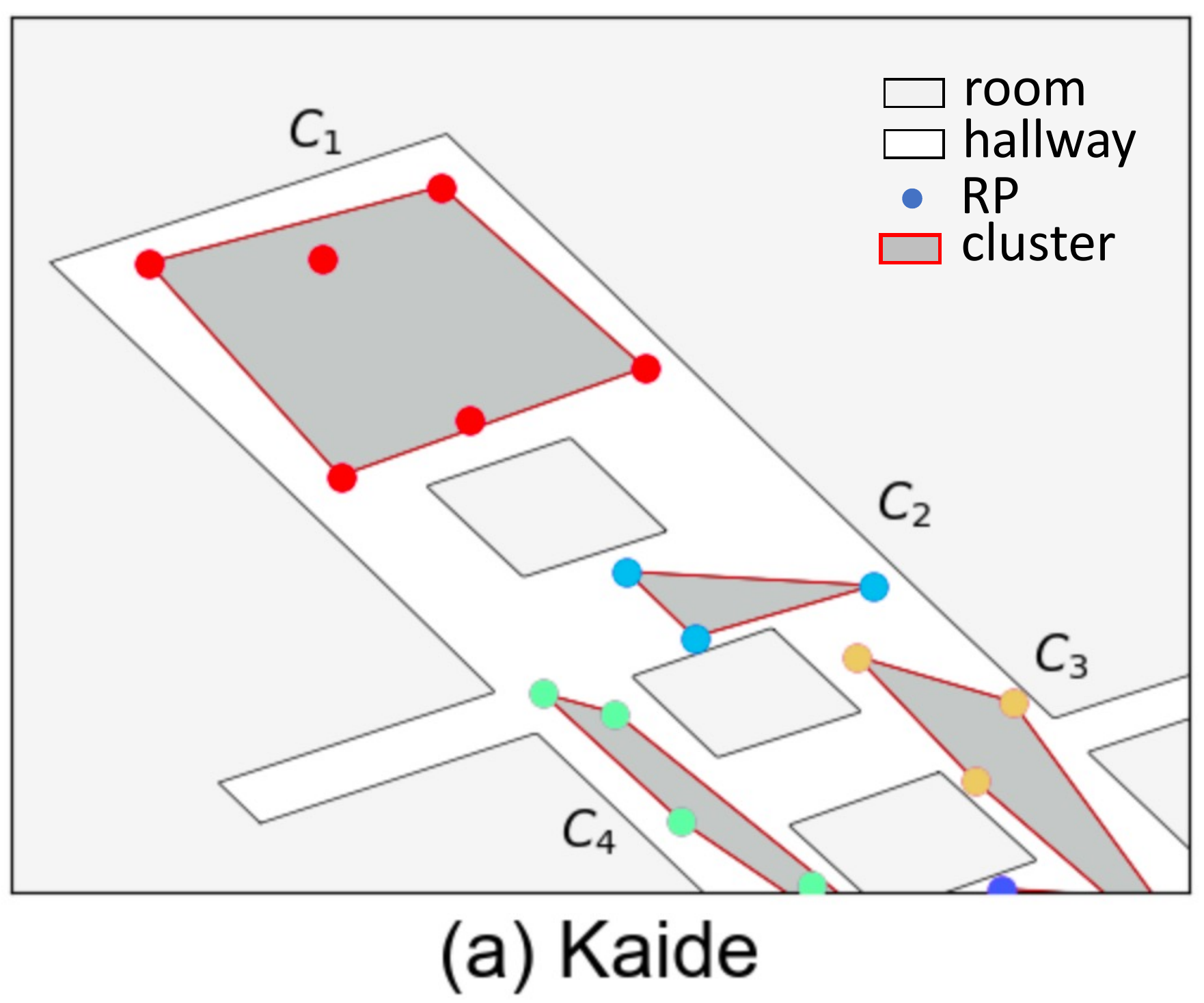}
    \end{minipage}
    \begin{minipage}[t]{0.236\textwidth}
    \centering
    \includegraphics[width=\textwidth]{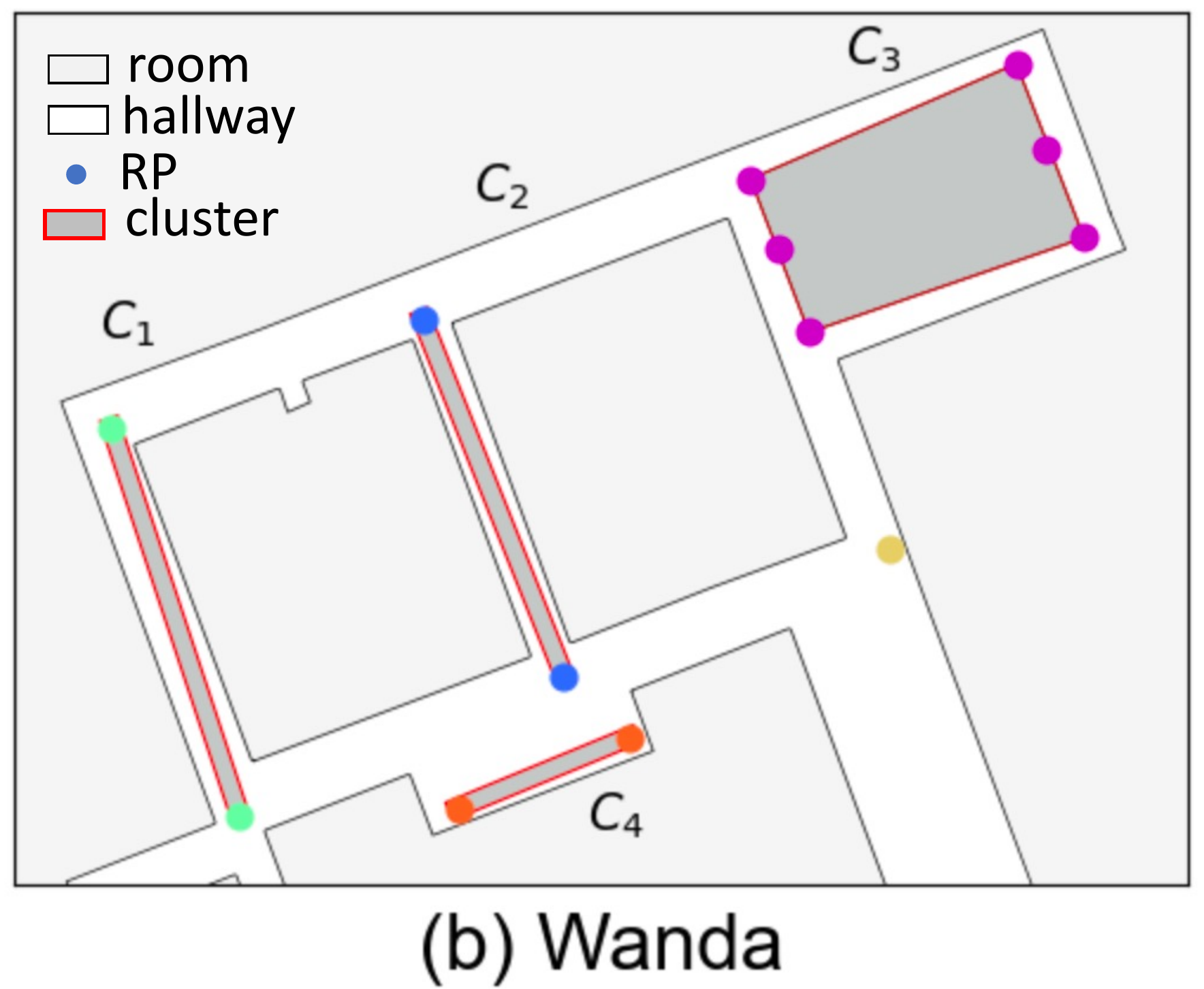}
    \end{minipage}
    \caption{Result of \TAC{}.}\label{fig:thac}
\end{figure}

Results of \TAC{} for the settings in Fig.~\ref{fig:akm} are visualized in Fig.~\ref{fig:thac}.
Each abnormal cluster in Fig.~\ref{fig:akm} is divided into  smaller clusters, each  spanning an open area.
\vspace{5pt}
\vspace{5pt}

\section{Data Imputer}
\label{sec:imputation}

After the  differentiation of missing RSSIs, the data imputer first replaces all identified \MNAR{}s with $-100$ dBm and changes their corresponding $-1$ in the mask matrix $\mathbf{M}$ to $1$.
The amended matrix, denoted as $\mathbf{M}'$, contains $0$s only for \MAR{}s and $1$s for \MNAR{}s and observed RSSIs.
 
Subsequently, the data imputer imputes \MAR{}s and RP \NULL{}s jointly using a sequential neural network.
The intuition is that radio map records on the same survey path are temporally correlated and the fingerprint and RP in one record are also correlated.
To capture the correlations among sequential records and in each radio map record, we propose a Bi-directional Sequence-to-Sequence Imputation Model (\BISIM{}).

\subsection{\BISIM{} Architecture}
\label{ssec:architecture}

\begin{figure*}[!ht]
\centering
\begin{minipage}[t]{0.9\textwidth}
\centering
\includegraphics[width=1\textwidth]{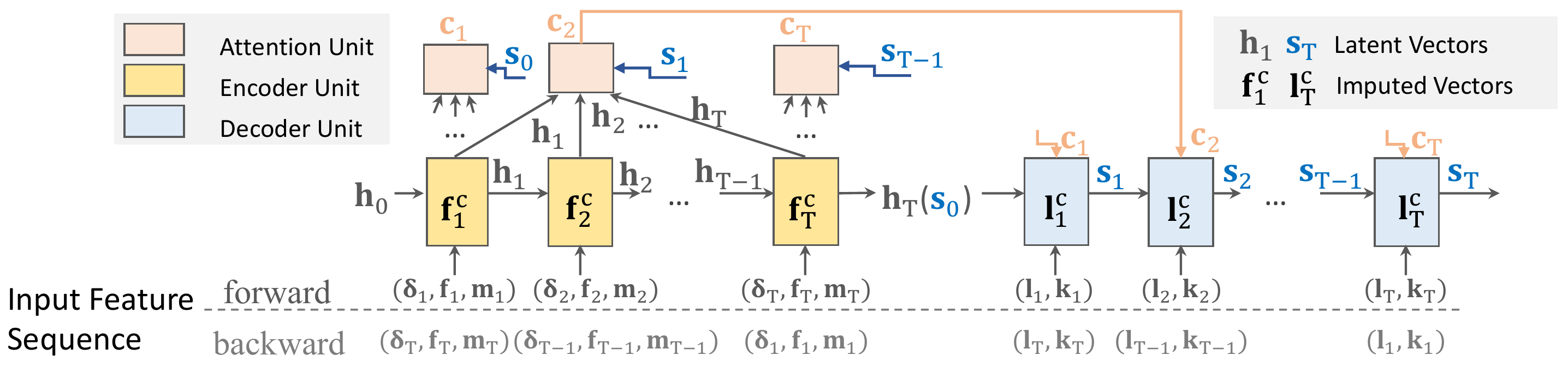}
\caption{The encoder-decoder architecture of \BISIM{}.}\label{fig:bisim_architecture}
\end{minipage}
\vspace{10pt}
\begin{minipage}[t]{0.3\textwidth}
\centering
\includegraphics[width=1\textwidth]{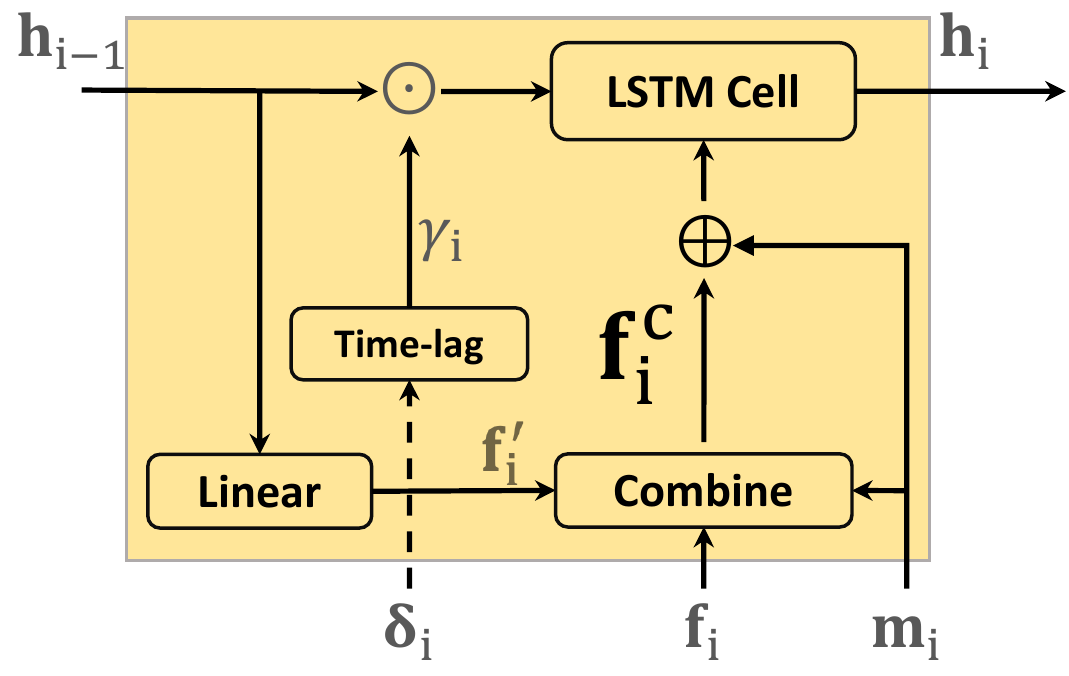}
\caption{The $i$th encoder unit. }\label{fig:encoder}
\end{minipage}
\begin{minipage}[t]{0.3\textwidth}
\centering
\includegraphics[width=1\textwidth]{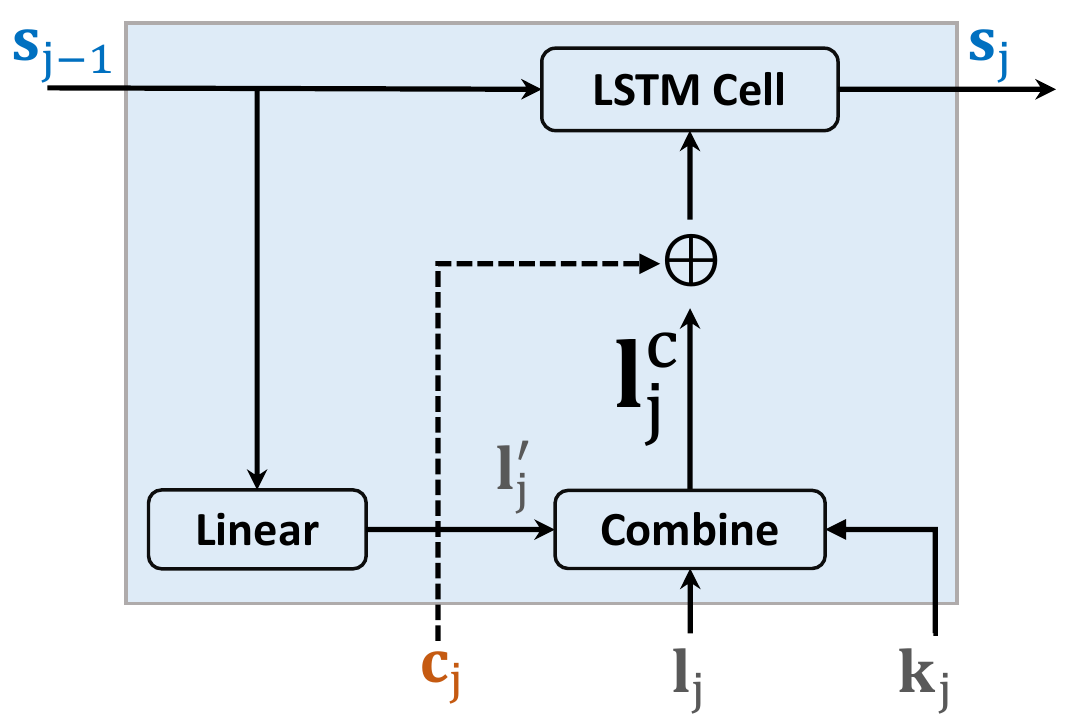}
\caption{The $j$th decoder unit.}\label{fig:decoder}
\end{minipage}
\begin{minipage}[t]{0.3\textwidth}
\centering
\includegraphics[width=0.8\textwidth]{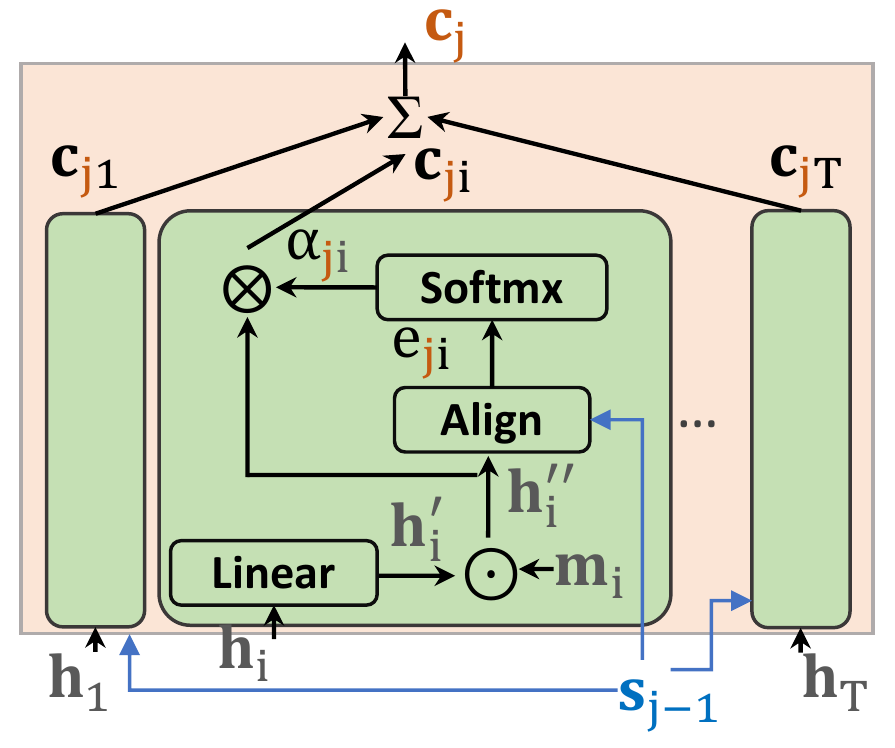}
\caption{The $j$th attention unit. }\label{fig:attention}
\end{minipage}
\end{figure*}

The \BISIM{} architecture is shown in Fig.~\ref{fig:bisim_architecture}.
The encoder-decoder~\cite{cho2014properties} architecture enables \BISIM{} to handle heterogeneous input data such that  fingerprint  and RP data sequences can be fed into the encoder and decoder units, respectively.
In Fig.~\ref{fig:bisim_architecture}, the tail of the encoders (the yellow part) is connected to the head of the decoders (the blue part) via a hidden vector $\mathbf{h}_T = \mathbf{s}_0$, meaning that the fingerprint sequence can decode the underlying RP sequence.
Note that conventional RNN-based imputation models~\cite{che2018recurrent,   yoon2017multi,cao2018brits, miao2021generative, cini2021multivariate} can only handle homogeneous data sequences and thus fall short in our setting. 

In general, \BISIM{} receives a sequence of $T$ radio map records on a survey path as input and outputs a corresponding sequence of $T$ imputed records.
Its data flow is as follows.

First, the features of the $i$th ($1 \leq i \leq T$) fingerprint in the sequence is fed to an \textbf{encoder unit}.
The input feature consists of three components $(\boldsymbol\delta_i, \mathbf{f}_i, \mathbf{m}_i)$, to be detailed in Section~\ref{ssec:bisim_input}.
The $i$th encoder unit generates an imputed vector $\mathbf{f}^{c}_i$ as well as a latent vector $\mathbf{h}_i$ to be passed to the next encoder unit. 
The initial latent vector $\mathbf{h}_0$ is randomized.

Second, the features of the $j$th ($1 \leq j \leq T$) RP in the sequence is fed to a \textbf{decoder unit}.
As also to be introduced in Section~\ref{ssec:bisim_input}, its input consists of two components $(\mathbf{l}_j, \mathbf{k}_j)$.
The $j$th decoder unit transforms the input features into an imputed RP vector $\mathbf{l}^c_j$ by utilizing the latent vector $\mathbf{s}_{j-1}$ from its preceding decoder unit, and it generates $\mathbf{s}_{j}$ that will be passed to the next decoder unit.

As the latent vector $\mathbf{s}_0$ ($\mathbf{h}_T$) is learned from the fingerprint sequence as a whole, we introduce a sparsity-friendly attention mechanism to make the decoder unit aware of on which parts of the fingerprint sequence to focus.
The $j$th \textbf{attention unit} (in pink in Fig.~\ref{fig:bisim_architecture}) receives the latent vectors $\{ \mathbf{h}_{1}, \ldots, \mathbf{h}_{T}\}$ from all encoder units and the latent vector $\mathbf{s}_{j-1}$ from the $(j-1)$th decoder unit and then generates a context vector $\mathbf{c}_j$ that is passed to the $j$th decoder unit for generating $\mathbf{l}^c_j$.

The internals of \BISIM{}, including the encoder unit, decoder unit, and attention unit, are detailed in Section~\ref{ssec:bisim_internals}.
Above, we covered the encoding-decoding process in the forward direction. 
Indeed, we also  capture the backward dependencies of the feature sequences. As shown at the bottom of Fig.~\ref{fig:bisim_architecture}, we feed the feature sequences backwards  to obtain another set of imputed vectors.
Our loss function takes into account the imputed vectors obtained from both forward and backward inputs, to be covered in Section~\ref{ssec:bisim_loss}.

\subsection{Input Feature Preparation}
\label{ssec:bisim_input}


\ptitle{Fingerprint Input Feature}
Given a fingerprint $\mathbf{f}_i$, the corresponding row in the mask matrix $\mathbf{M}'$ is retrieved as $\mathbf{m}_i$.
The mask vector $\mathbf{m}_i$ records which AP values of $\mathbf{f}_i$ are \NULL{}s.
In the encoding stack, each unit's encoding depends on the latent vector from the previous unit.
Intuitively, a latent vector from a more distant time should exert less influence on the current unit. To reflect this time decay effect on encoding, we introduce a \emph{time-lag vector}~\cite{cao2018brits, miao2021generative} $\boldsymbol\delta_i = \langle \delta_i^1, \ldots, \delta_i^j, \ldots, \delta_i^D \rangle$ for each input fingerprint $\mathbf{f}_i$, where
\begin{equation}\label{equation:time_lag_vectors}
\begin{aligned}
\delta_i^j & = 
\begin{cases}
0 & \text{if~} i=1 \\
t_i-t_{i-1} & \text{if~} i>1 \land \mathbf{m}[i-1,j]=1\\
\delta_{i-1}^j + (t_i - t_{i-1}) & \text{if~} i>1 \land \mathbf{m}[i-1,j] = 0\\
\end{cases}
\end{aligned}
\end{equation}

In Eq.~\ref{equation:time_lag_vectors}, each time-lag vector value $\delta^j_i$ for the first encoder unit is set to $0$ by default. For other units, we differentiate two cases. If the previous observation is not \NULL{} (i.e., $\mathbf{m}[i-1,j] = 1$), the value is simply the difference between the current time and the previous time, i.e., $t_i-t_{i-1}$. Otherwise, the value is 
the sum of $(t_i - t_{i-1})$ and $\delta_{i-1}^j$ (the value of the previous time). 
Note that only observed values from the previous time affect the current encoder unit.
In this sense, $\delta^j_i$ in Eq.~\ref{equation:time_lag_vectors} keeps track of the difference between the current time and the last observation's time. 


\ptitle{RP Input Feature}
Given a RP $\mathbf{l}_j$, we generate a mask vector $\mathbf{k}_j \in \{ 0,1 \}^2$ as follows.
If $\mathbf{l}_j$ is not \NULL{} then $\mathbf{k}_j = \langle 1, 1 \rangle$; otherwise, $\mathbf{k}_j = \langle 0, 0 \rangle$.
We have generated a similar time-lag vector for $\mathbf{l}_j$ as the decoder input.
However, ablation studies in Section~\ref{ssec:evaluation_data_imputation} show such extra decoder input brings about no gains.
As time decay has been captured by encoder units, a more complex structure may degrade model generalizability.

\begin{example}
Table~\ref{tab:input_feature} shows mask vectors $\mathbf{m}_1$ to $\mathbf{m}_5$ and $\mathbf{k}_1$ to $\mathbf{k}_5$ for Table~\ref{tab:incmp_radio_map}.
Fingerprints' time-lag vectors are generated as follows.
According to Eq.~\ref{equation:time_lag_vectors}, $\boldsymbol\delta_1$ is simply $\langle 0,0,0,0,0 \rangle$.
For $\mathbf{f}_2$ of  time $t_3 = 3$ in Table~\ref{tab:incmp_radio_map}, the values $\delta^1_2$ to $\delta^3_2$ all equal to $t_3 - t_1 = 3$; the value $\delta^4_2$ equals to $\delta^4_1 + (t_3 - t_1) = 3$ as $\delta^4_1 = 0$, and $\delta^5_2 = 3$ follows a similar computation as $\delta^4_2$.
For $\mathbf{f}_2$ of time $t_4 = 8$ in Table~\ref{tab:incmp_radio_map}, $\delta^1_3 = t_4 - t_3 = 8 - 3 = 5$, whereas $\delta^2_3 = \delta^1_2 + (t_4 - t_3) = 3 + (8 - 3) = 8$.
The subsequent computations are performed similarly.
\end{example}

\begin{table}[]
\centering
    \caption{Input Features for \BISIM{}}
    \label{tab:input_feature}
    \footnotesize
\begin{tabular}{|l|>{\columncolor[gray]{0.8}}l|l|l|l|l|l|>{\columncolor[gray]{0.8}}l|l|l|}
\hline
& & $r_1$ & $r_2$ & $r_3$ & $r_4$ & $r_5$ &    & $x$ & $y$ \\ \hline \hline
\multirow{5}{*}[0ex]{\rotatebox[origin=c]{90}{mask vec.}} & $\mathbf{m}_1$ & 1  & 1  & 1  & 0  & 0  & $\mathbf{k}_1$ & 1 & 1 \\ \cline{2-10}
& $\mathbf{m}_2$ & 1  & 0  & 1  & 0  & 0  & $\mathbf{k}_2$ & 0 & 0 \\ \cline{2-10}
& $\mathbf{m}_3$ & 0  & 0  & 1  & 1  & 0  & $\mathbf{k}_3$ & 1 & 1 \\ \cline{2-10}
& $\mathbf{m}_4$ & 1  & 1  & 0  & 0  & 1  & $\mathbf{k}_4$ & 0 & 0 \\ \cline{2-10}
& $\mathbf{m}_5$ & 0  & 0  & 0  & 0  & 0  & $\mathbf{k}_5$ & 1 & 1 \\ \hline 
\hline
\multirow{5}{*}[0ex]{\rotatebox[origin=c]{90}{time-lag vec.}} & $\boldsymbol\delta_1$ & 0  & 0  & 0  & 0  & 0  & & &  \\ \cline{2-10}
& $\boldsymbol\delta_2$ & 3  & 3  & 3  & 3  & 3  & & &  \\ \cline{2-10}
& $\boldsymbol\delta_3$ & 5  & 8  & 5  & 8  & 8  & & &  \\ \cline{2-10}
& $\boldsymbol\delta_4$ & 9  & 12  & 4  & 4  & 12  & & &  \\ \cline{2-10}
& $\boldsymbol\delta_5$ & 4  & 4  & 8  & 8  & 4  & & &  \\ \hline
\end{tabular}
\vspace*{-8pt}
\end{table}

\subsection{Internals of \BISIM{}}
\label{ssec:bisim_internals}

Unlike traditional encoder-decoder models, \BISIM{} must handle the \NULL{}s in the network units.
This is achieved by including the mask vectors $\mathbf{m}_i$ and $\mathbf{k}_j$ in the computation.
Taking the forward feature input as an example, we elaborate on each type of unit as follows.

\ptitle{Encoder Unit}
Fig.~\ref{fig:encoder} shows the $i$th encoder unit's internals. It takes $\mathbf{f}_i$, $\mathbf{m}_i$, $\boldsymbol\delta_i$, and the previous latent vector $\mathbf{h}_{i-1}$ as input, and it generates an intermediate imputed vector $\mathbf{f}^c_i$ and the current latent vector $\mathbf{h}_i$.
The formulas are given below.
\begin{equation}
\label{equ:encoder_linear_fp}
\mathbf{f}'_i = \mathbf{W}_f \mathbf{h}_{i-1} + \mathbf{b}_f
\end{equation}
\begin{equation}
\label{equ:encoder_complement_fp}
\mathbf{f}^c_i = \mathbf{m}_i \odot \mathbf{f}_i + (\mathbf{1}-\mathbf{m}_i)\odot \mathbf{f}'_i
\end{equation}
\begin{equation}
\label{equ:encoder_decay_h}
 \boldsymbol{\gamma}_i = \exp\big(-\max(0, \mathbf{W}_{\gamma}\boldsymbol{\delta}_i + \mathbf{b}_{\gamma})\big)
\end{equation}
\begin{equation}
\label{equ:encoder_feed_forward}
\mathbf{h}_i = \sigma \big( \mathbf{W}_h (\mathbf{h}_{i-1}\odot \boldsymbol{\gamma}_i) + \mathbf{U}_h (\mathbf{f}^c_i \oplus \mathbf{m}_i) + \mathbf{b}_h \big)
\end{equation}

Above,  matrices $\mathbf{W}_{*}$ and $\mathbf{U}_{*}$ and vectors $\mathbf{b}_{*}$ in the network units are learnable parameters.
Eq.~\ref{equ:encoder_linear_fp} is a linear operator that maps the previous latent vector $\mathbf{h}_{i-1}$ to an estimated fingerprint vector $\mathbf{f}'_i$.
Eq.~\ref{equ:encoder_complement_fp} is a combination operator that replaces the missing value in the fingerprint $\mathbf{f}_i$ with the corresponding values in the estimated fingerprint $\mathbf{f}'_i$.
It performs an element-wise product (i.e., $\odot$) of the fingerprint vectors and the mask vector $\mathbf{m}_i$.
The resulting complemented vector $\mathbf{f}^c_i$  forms the imputation result for $\mathbf{f}_i$ (see Eq.~\ref{equation:bisim_output}).
Eq.~\ref{equ:encoder_decay_h} generates a scalar \emph{temporal decay factor} $\gamma_i$ based on the time-lag vector $\boldsymbol\delta_i$.
Generally speaking, a larger $\boldsymbol\delta_i$ leads to a smaller $\gamma_i$, capturing that the effect of a past observation is reduced if the observation is temporally distant.
Finally, the temporal decay factor $\gamma_i$ is applied to $\mathbf{h}_{i-1}$, and the result is passed to a standard LSTM cell along with the imputed fingerprint $\mathbf{f}^c_i$ concatenated with $\mathbf{m}_i$. The LSTM cell's computation is formalized in Eq.~\ref{equ:encoder_feed_forward}, where $\sigma(\cdot)$ is the sigmoid function and $\oplus$ is the concatenation operator.

\ptitle{Decoder Unit}
Shown in Fig.~\ref{fig:decoder}, the internal of a decoder unit is similar to that of an encoder unit, except that no time-lag vector is used.
The formulas are given below.
In particular, the latent vector $\mathbf{s}_{j-1}$ is mapped to an estimated RP vector $\mathbf{l}'_j$ through a linear operator (Eq.~\ref{equ:dec_reconstru}).
Then, $\mathbf{l}'_j$ is used to replace the \NULL{} RP vector $\mathbf{l}_j$ in a combination operation (Eq.~\ref{equ:combine}).
Finally, the  concatenation of the resulting imputed vector $\mathbf{l}^c_j$ and  the context vector $\mathbf{c}_j$ is passed to an LSTM cell along with the latent vector $\mathbf{s}_{j-1}$.
The LSTM cell in Eq.~\ref{equ:dec_propagate} generates the latent vector $\mathbf{s}_{j}$ for the next unit. 
%
\begin{equation}
\label{equ:dec_reconstru}
\mathbf{l}'_j = \mathbf{W}_l \mathbf{s}_{j-1} + \mathbf{b}_l
\end{equation}
\begin{equation}
\label{equ:combine}\mathbf{l}_j^c = \mathbf{k}_j \odot \mathbf{l}_j + (\mathbf{1} - \mathbf{k}_j ) \odot \mathbf{l}'_j
\end{equation}
\begin{equation}
\label{equ:dec_propagate}
\mathbf{s}_j = \sigma \big(\mathbf{W}_s\mathbf{s}_{j-1} + \mathbf{U}_s(\mathbf{l}_j^c \oplus \mathbf{c}_j) + \mathbf{b}_s \big)
\end{equation}

\ptitle{Attention Unit}
As shown in Fig.~\ref{fig:attention}, the $j$th attention unit generates a context vector $\mathbf{c}_j$ to help the $j$th decoder selectively retrieve information from the fingerprint sequence in decoding the corresponding RP vector.
We employ the Bahdanau attention mechanism~\cite{bahdanau2014neural}, which can dynamically capture the relationship between the current decoding moment and each past encoding moment and then assign higher attention (i.e., weights) to the more related encoding moments.
However, the original Bahdanau attention does not consider the incompleteness in the input of an encoder unit, which may involve noise in the resulting latent vector.
To avoid this, we design a sparsity-friendly variant of the  Bahdanau attention, by  allowing only observed values' latent vectors to participate in the computation. Specifically in Eq.~\ref{equ:att_filter_ho}, we  transform  each latent vector  $\mathbf{h}_i$ linearly to $\mathbf{h}'_i$ and retain only the observed part of $\mathbf{h}'_i$ by performing an element-wise product of $\mathbf{h}'_i$ and $\mathbf{m}_i$.
%

\begin{equation}
\label{equ:att_filter_ho}
\mathbf{h}'_i = \mathbf{W}_a \mathbf{h}_i + \mathbf{b}_a; \;\;\;  \mathbf{h}''_i = \mathbf{h}'_i \odot \mathbf{m}_i
\end{equation}
\begin{equation}
\label{equ:att_score_ori}
e_{ji} = \text{MLP}(\mathbf{s}_{j-1}, \mathbf{h}''_i)
\end{equation}
\begin{equation}
\label{equ:att_score_softmax}
\alpha_{ji} = \exp(e_{ji})/\sum\nolimits_{k=1}^T\exp(e_{jk})
\end{equation}
\begin{equation}
\label{equ:att_context}
\mathbf{c}_j = \sum\nolimits_{i=1}^T \mathbf{c}_{ji}; \;\;\; \mathbf{c}_{ji} =  \alpha_{ji}\mathbf{h}''_i
\end{equation}

Next, Eq.~\ref{equ:att_score_ori} --~\ref{equ:att_context} use the original Bahdanau attention~\cite{bahdanau2014neural}.
In particular, Eq.~\ref{equ:att_score_ori} implements an alignment function that aligns $\mathbf{s}_{j-1}$ and $\mathbf{h}''_i$ into an energy factor $e_{ji}$ based on a Multilayer Perceptron (MLP).
The energy factor reflects the importance of the encoder's latent vector $\mathbf{h}''_i$ with respect to the decoder's latent vector $\mathbf{s}_{j-1}$ in generating $\mathbf{s}_j$, the next decoder's latent vector.
Afterwards, $e_{ji}$ is normalized into a weight $\alpha_{ji}$ by a softmax function, 
 in Eq.~\ref{equ:att_score_softmax}. 
With such weights, we calculate the context vector $\mathbf{c}_j$ as a weighted sum of all $\mathbf{h}''_i$s,  in Eq.~\ref{equ:att_context}. 

\subsection{Output and Loss Function}
\label{ssec:bisim_loss}

Recall that we generate two  pairs of imputed vectors, i.e., $\mathbf{f}^c_{i,\succ}$ and $\mathbf{l}^c_{i,\succ}$ for forward input features, and $\mathbf{f}^c_{i,\prec}$ and $\mathbf{l}^c_{i,\prec}$ for backward input features. We average the vectors from both directions to get the final output. Formally, we have:  \begin{equation}\label{equation:bisim_output}
\hat{\mathbf{l}}_i = (\mathbf{l}^c_{i,\succ} + \mathbf{l}^c_{i,\prec})/2; \,\,\,\,\,\, \hat{\mathbf{f}}_i = (\mathbf{f}^c_{i,\succ} + \mathbf{f}^c_{i,\prec})/2
\end{equation}
{As we lack ground-truth of the imputed results in model training, we base our loss function on the reconstruction errors between the observed values in the radio map and the corresponding values predicted by the model.
Intuitively, if the model makes predictions close to the original observed values, the model is likely to impute missing values reliably~\cite{cao2018brits}.}
The overall loss $\mathcal{L}^o$ of \BISIM{} is defined as follows.
\begin{equation*}\label{equation:loss}
\begin{aligned}
& \mathcal{L}^o = \mathcal{L}^\mathit{forward} +  \mathcal{L}^\mathit{backward} + \mathcal{L}^\mathit{cross}, \text{where}~\\
& \mathcal{L}^\mathit{forward} =  1/T\cdot\sum\nolimits_{i=1}^{T} \big( \mathcal{L}(\mathbf{f}'_{i,\succ}, \mathbf{f}_{i,\succ}, \mathbf{m}_i) + \mathcal{L}(\mathbf{l}'_{i,\succ}, \mathbf{l}_{i,\succ}, \mathbf{k}_i) \big) \\
& \mathcal{L}^\mathit{backward} =  1/T\cdot\sum\nolimits_{i=1}^{T} \big( \mathcal{L}(\mathbf{f}'_{i,\prec}, \mathbf{f}_{i,\prec}, \mathbf{m}_i) + \mathcal{L}(\mathbf{l}'_{i,\prec}, \mathbf{l}_{i,\prec}, \mathbf{k}_i) \big) \\
& \mathcal{L}^\mathit{cross} =  1/T\cdot\sum\nolimits_{i=1}^{T} \big( \mathcal{L}(\mathbf{f}'_{i,\succ}, \mathbf{f}'_{i,\prec}, \mathbf{m}_i) + \mathcal{L}(\mathbf{l}'_{i,\succ}, \mathbf{l}'_{i,\prec}, \mathbf{k}_i) \big) \\
& \mathcal{L}(\mathbf{a}, \mathbf{a}', \mathbf{mask}) = \text{MSE}(\mathbf{mask} \odot \mathbf{a}, \mathbf{mask} \odot \mathbf{a}')
\end{aligned}
\end{equation*}
Above,  $\mathbf{f}_{i,\succ}$ and $\mathbf{l}_{i,\succ}$ (resp. $\mathbf{f}_{i,\prec}$ and $\mathbf{l}_{i,\prec}$) are the forward (resp. backward) input features.
The overall loss $\mathcal{L}^o$ consists of three terms. The forward loss $\mathcal{L}^\mathit{forward}$ captures the reconstruction error of the forward imputation results. The backward loss $\mathcal{L}^\mathit{backward}$ captures the reconstruction error of the backward imputation results. The cross loss $\mathcal{L}^\mathit{cross}$ captures the closeness between each pair of  forward  and backward imputation results.
To measure reconstruction errors, we use the predicted vector (e.g., $\mathbf{f}'_{i,\succ}$ in Eq.~\ref{equ:encoder_linear_fp}) instead of the final imputation result (e.g., $\mathbf{f}^c_{i,\succ}$) because the observed part of the final imputation result comes directly from the input feature (e.g., $\mathbf{f}_{i,\succ}$).
{ The function $\mathcal{L}(\mathbf{a}, \mathbf{a}', \mathbf{mask})$ measures the MSE (mean square error) between the \emph{observed} parts of the input vectors $\mathbf{a}$ and $\mathbf{a}'$,
where $\mathbf{mask}$ is a mask vector for retaining the original observed values in the input vectors. In particular, $\mathbf{mask}$ is $\mathbf{m}$ and $\mathbf{k}$ for fingerprints and RPs, respectively.}

\section{Experimental Studies}
\label{sec:exp}

\subsection{Experimental Settings}
\label{ssec:settings}

All algorithms are coded in Python 3.8 and run on a Linux server with 3.60 GHz Intel Core i9 CPU and NVIDIA RTX 3080 GPU with 12 GB memory. All neural network models are implemented using PyTorch 1.6 and trained on the GPU.
The code, datasets, and tuning details are available online~\cite{codeRepo}.

\ptitle{Datasets and Real Indoor Venues}
We use  real-world indoor positioning datasets~\cite{dataset} published by Microsoft Research, which encompass  walking survey records, building topological information, and online testing data collected from 
shopping malls in China. For our studies, we randomly pick  two malls: Kaide Mall and Wanda Square as Wi-Fi fingerprinting scenarios. In addition, to gain insights into  the effectiveness of our proposals in other application scenarios and indoor venues, we conducted additional experiments using Bluetooth fingerprinting data from a different indoor venue named \SVI{}.
%
For radio map creation, the parameter $\epsilon$ is set to $1$ second for both venues. 
The characteristics of the venues and radio maps are given in Table~\ref{tab:dataset_des}.~\SV{} features a larger radio map with a higher fingerprint dimensionality and more fingerprints, whereas \SIV{} features a higher RP density. Note that the APs in Longhu are Bluetooth-based instead of Wi-Fi based.
%
%

\vspace{10pt}
\begin{table}[!htbp]
    \centering
    \footnotesize
    {\setlength\tabcolsep{5pt}
    \caption{Statistics of Venues and Created Radio Maps}
    \label{tab:dataset_des}
    \begin{tabular}{l|ccc}
    \toprule
    {Venue} & \SIV{} & \SV{} & \SVI{}  \\ 
    \midrule
    Floor Area (m$^2$) & 3225.7 & 4458.5 & 6504.1  \\
    RP density (per 100 m$^2$) & 3.53 & 2.65 & 3.11   \\
    \# of fingerprints & 894 & 4104 & 4617 \\
    \# of RPs & 114 & 118 &202 \\
    \# of APs (i.e., \# of fingerprint dimensions) & 671 & 929 & 330 \\
    \bottomrule
    \end{tabular}}
\end{table}
\vspace{10pt}

\ptitle{Evaluation Controls}
To evaluate our overall solution framework with an \MNAR{}/\MAR{} differentiator \textsf{A} and a data imputer \textsf{B}, we employ an online location estimation algorithm \textsf{C} as follows.
Given an original radio map, we select 10\% of the records with observed RPs as testing data and use the RPs as ground-truth locations for evaluation.
Modules \textsf{A} and \textsf{B} are combined to impute both testing data and the rest radio map records. After that, the remaining records form a radio map used by \textsf{C} to estimate the locations on the testing data\footnote{We also apply imputation to the (online) fingerprints in the test data. Usually, complete online fingerprints are obtained by using techniques unavailable or unaffordable for walking surveys~\cite{khalajmehrabadi2017modern}.}.

Given different combinations of \textsf{A}, \textsf{B}, and \textsf{C}, we use the method of control variates in the evaluations.
In Section~\ref{ssec:evaluation_differentiation}, we compare different differentiators (\textsf{A}), fixing \textsf{B} to \BISIM{} and \textsf{C} to \WKNN{}. 
\BISIM{} and \WKNN{} together perform best across different differentiators, to be shown in Section~\ref{ssec:evaluation_data_imputation}, where we compare different data imputers (\textsf{B}) across different combinations of \textsf{A} and \textsf{C}.

\subsection{Evaluation of Differentiators}
\label{ssec:evaluation_differentiation}
\subsubsection{Setting}

\fptitle{Methods}
Based on Algorithm~\ref{alg:differentiation}, we evaluate three differentiators using different clustering methods\footnote{We omit the inferior results of other clustering methods like DBSCAN.}, namely our \AKM{} and \TAC{}, and  $K$-means based on the elbow method for $K$ selection~\cite{kargar2021predict} (denoted as \ELKM{}).
For \AKM{} and \ELKM{} that decide $K$ through iterations, we set $K$'s upper-bound $U$ to {$200$}. 
To sample ground-truth sets in \AKM{}, we fix the number of sampled \MNAR{}s ($6960$ for \SIV{} and $9612$ for \SV{}) and take the proportion $\gamma = \frac{\text{\#(\MNAR{}s)}}{\text{\#(\MAR{}s)}}$ from the list $\Gamma = (1, 2, \ldots, 20)$.
The proportion starts from 1 as there should be more \MNAR{}s than \MAR{}s (i.e., random events) in practice.
%
We also implement two baselines without differentiation: \textsf{MAR-only} treats all missing RSSIs as \MAR{}s, and \textsf{MNAR-only} treats all as \MNAR{}s.

\ptitle{Parameters}
First, we examine how differentiators are affected by the 
sparsity
of input radio maps.
Specifically, we introduce a \emph{removal ratio} $\alpha \in \{0, 5, 10, 15, 20\}$\% 
such that a fraction $\alpha$ of RSSIs are randomly selected and nullified in an original radio map. As a result, the input radio map has $\{85.6, 86.3, 87.0, 87.7, 88.4\}$\% missing RSSIs for \SIV{}, and $\{93.1, 93.4, 93.7, 94.0, 94.3\}$\% missing RSSIs for \SV{}. We test the performance of differentiators under such high missing rates of RSSIs in the input radio map.

%
Further, we test the effect of the fraction threshold $\eta$ in Algorithm~\ref{alg:differentiation} on the differentiators by varying it in $\{0, 0.1, 0.2, 0.3\}$.  
By default, we set $\alpha = 0$ and $\eta = 0.1$.
In each test, we vary one parameter and set the others to  default.

\ptitle{Metrics}
We measure the \textbf{average positioning error} (\textbf{\APE{}}) between all estimated locations and their ground-truth locations.
Note that we do not evaluate the differentiators using the \DA{} metric. As \DA{} is utilized in \AKM{} (and  not in the other methods), this could lead to an unfair comparison.

\subsubsection{Results}

\begin{figure}[h]
\centering
\includegraphics[width=\columnwidth]{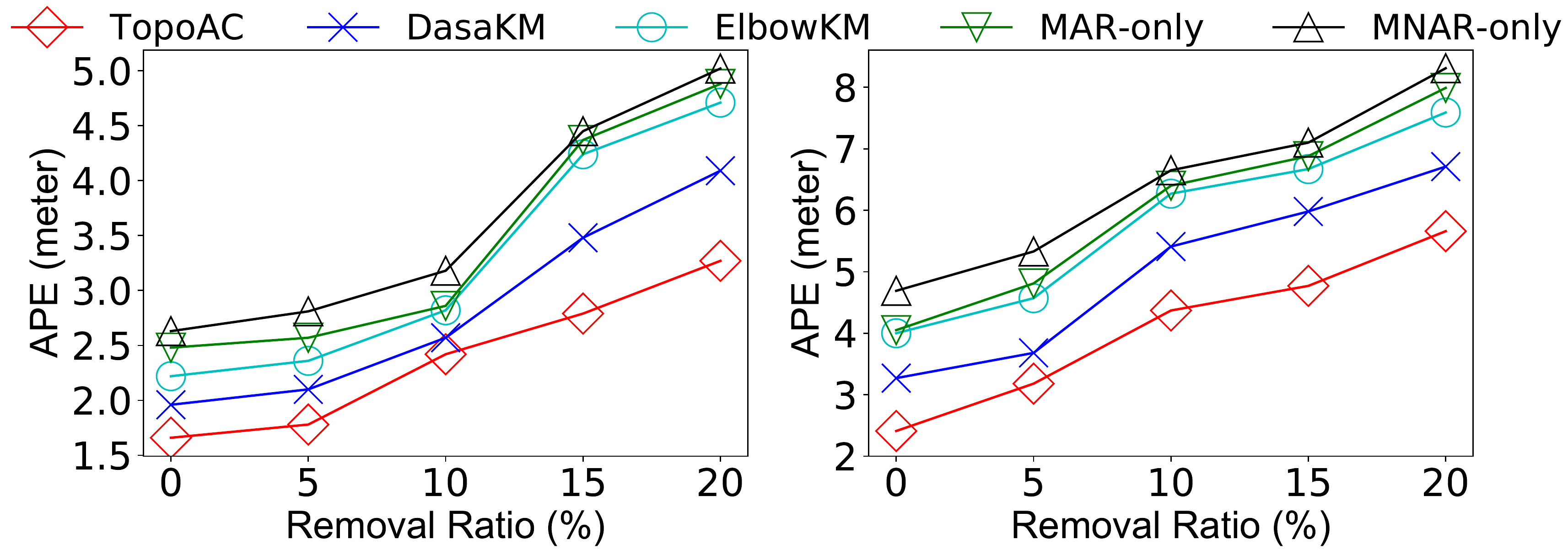}\\
\vspace*{-4pt}
{\small\ \ \ \ \ (a) \SIV{} \ \ \ \  \ \ \ \ \ \ \  \ \ \ \ \ \ \ \ \ \ \ \ \  (b) \SV{} }
\caption{The removal ratio $\alpha$ vs. APE.}
\label{fig:ape_vary_eta}
\end{figure}

\ptitle{Effect of Removal Ratio $\alpha$}
The \APE{} results for different removal ratios
are reported in Fig.~\ref{fig:ape_vary_eta}. 
All methods 
 are affected negatively by a larger $\alpha$ since more observed values are removed, which reduces the final positioning accuracy. Also, the three differentiator methods consistently outperform \textsf{MAR-only} and \textsf{MNAR-only}, showing the significance of differentiation.
 By distinguishing \MNAR{}s and \MAR{}s  and imputing them differently, these methods reduce  bias that exists in \textsf{MAR-only} and \textsf{MNAR-only} methods which  treat  all the missing RSSIs as the same kind in the imputation.
%
\textsf{MAR-only} always outperforms \textsf{MNAR-only}---\textsf{MNAR-only} naively fills in all missing RSSI with $-100$ dBm, while  \textsf{MAR-only} employs the imputer to approach the true values of missing RSSIs.
Regarding the differentiators, \ELKM{} performs worse than \AKM{} and \TAC{},  and its performance degrades more rapidly.
Due to its inferiority, \ELKM{} is excluded from  evaluations of data imputers in Section~\ref{ssec:evaluation_data_imputation}.

Compared to \ELKM{}, \AKM{} improves the positioning accuracy by more than $0.3$ m in \SIV{} and by more than $0.4$ m in \SV{}. In practice, a positioning error of $0.3$ m is likely to localize a user mistakenly to another room behind a wall, thus impairing the quality of downstream services such as indoor navigation and contact tracing. 
Overall, the proposed differentiation accuracy (\DA{}) is shown to be effective and necessary for the $K$-means based missing RSSI differentiation.


Compared to \AKM{}, \TAC{} requires no brute-force $K$ search or \DA{} measurement, while achieving better \APE{} in all tests.
This shows the effectiveness of utilizing indoor topology in clustering.
However, in case topological information is unavailable, the proposed \AKM{} provides a useful, alternative method for missing RSSI differentiation.
%

\begin{figure}[h]
\vspace*{6pt}
\centering
\includegraphics[width=\columnwidth]{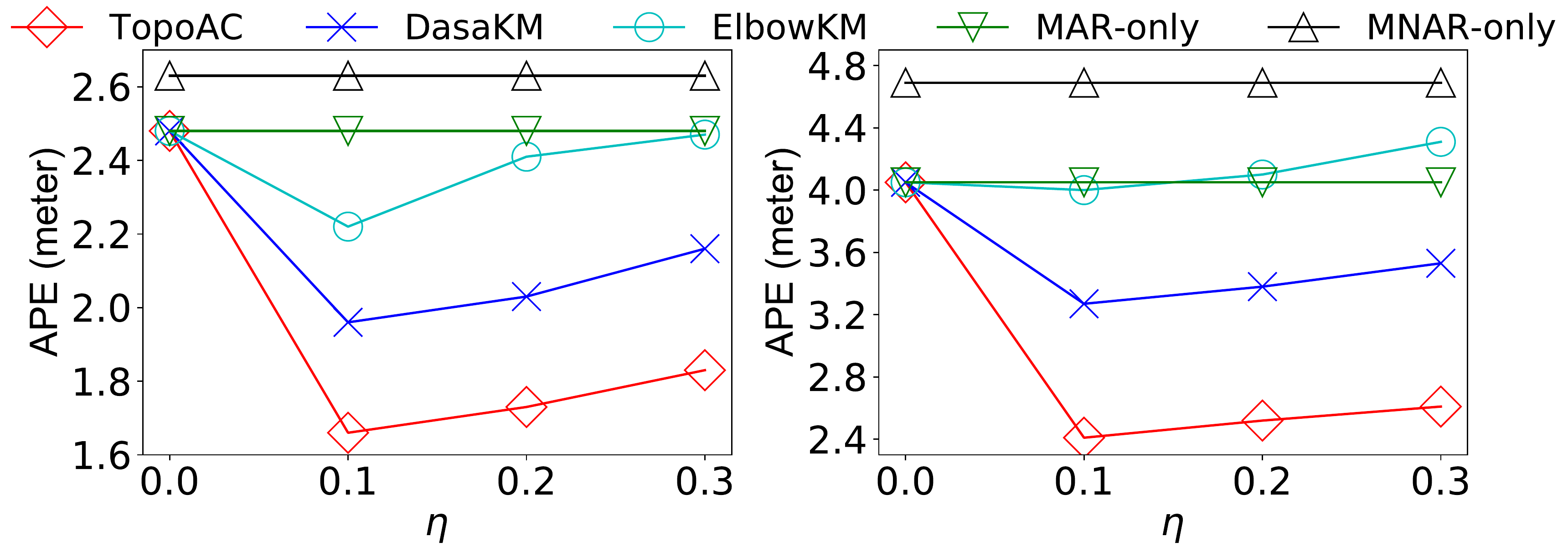}\\
\vspace*{-4pt}
{\small\  \ \ \ \ (a) \SIV{} \ \ \ \  \ \ \ \ \ \ \  \ \ \ \ \ \ \ \ \ \ \ \ \ (b) \SV{} }
\caption{The threshold $\eta$ vs. APE.}
\label{fig:ape_vary_thre}
\vspace*{-5pt}
\end{figure}

\ptitle{Effect of Fraction Threshold $\eta$} The  APE results are reported in Fig.~\ref{fig:ape_vary_thre}.   Threshold $\eta$ imposes requirements on the identification of \MAR{}s within a cluster. The setting  $\eta=0$ means that all differentiators consider missing RSSIs as \MAR{}s despite the clustering results; thus, they result in the same \APE{} as \textsf{MAR-only}.
As $\eta$ increases, the requirement for a missing RSSI to be judged as a \MAR{} becomes stricter, and thus previously incorrectly identified \MAR{}s are identified as \MNAR{}s, which initially improves the \APE{} for differentiators (cf. $\eta=0.1$). 
However, as $\eta$ increases further, more \MAR{}s are mistakenly recognized as \MNAR{}s, which leads to  worse \APE{} results. This can be highlighted by \ELKM{} (e.g., $\eta=0.3$ in Wanda), where the \APE{} is even higher than that of \textsf{MAR-only}. In contrast, \AKM{} and \TAC{} are more stable to the increasing $\eta$ thanks to their effectiveness in clustering similar AP profiles against identification errors. If $\eta$ goes up to 1, all three differentiators would have the same APE as \textsf{MNAR-only}, as all missing RSSIs are regarded as \MNAR{}s. Overall, \TAC{} outperforms the others, and $\eta=0.1$ is the best threshold for all differentiators.

\ptitle{Distribution of Differentiated Results}
Based on  \TAC{}'s differentiated results in the default setting, \MAR{}s account for 10.12\% of all missing RSSIs in \SIV{} and 7.06\% in \SV{}. Note that this is only an estimated result---as mentioned earlier, the real distribution is unknown.

\subsection{Evaluation of Data Imputers}
\label{ssec:evaluation_data_imputation}
\subsubsection{Setting}

\fptitle{Methods}
We implement two \BISIM{} variants: (1) \ABISIM{} combines \AKM{} and \BISIM{} and (2) \TBISIM{} combines \TAC{} and \BISIM{}.
In addition, we include the following data imputers: (3) \textbf{Case Deletion} (\CD{})~\cite{kaiser2014dealing} removes all radio map records with \texttt{null} RPs and uses $-100$ dBm for each missing RSSI; (4) \textbf{Linear Interpolation} (\LI{})~\cite{li2017turf} 
differs from \CD{} in that it 
 interpolates the missing RPs linearly based on their previously and subsequently observed RPs along a path. (5) \textbf{Semi-supervised Learning} (\SL{})~\cite{sorour2014joint} replaces RP interpolation in \LI{} by a semi-supervised model that utilizes records with observed RPs as samples for iterative inferencing of missing RPs; (6) \textbf{Multiple Imputation by Chained Equation} (\MICE{})~\cite{azur2011multiple} iteratively fills-in missing values of a column with other columns filled with their mean values by default; (7) \textbf{Matrix Factorization} (\MF{})~\cite{hastie2009elements} fills-in missing values in the radio map based on matrix completion; (8) \textbf{Bidirectional Recurrent Imputation for Time Series} (\BRITS{})~\cite{cao2018brits} captures time-series data dependencies based on an RNN for imputing \texttt{null} RSSIs with known RPs and uses the \LI{}\footnote{\BRITS{} cannot impute RSSIs and RPs jointly. The \BRITS{} variants with \CD{} and \SL{} to missing RPs achieve similar performance. We omit them here.} strategy to impute \texttt{null} RPs; (9) \textbf{Semi-Supervised Generative Adversarial Network} (\SSGAN{})~\cite{miao2021generative} is the state-of-the-art GAN model for multivariate time series imputation.

Note that \MICE{}, \MF{},  and \BRITS{} can also use \MAR{} results obtained by either \AKM{} or \TAC{}. These imputation methods achieve better performance when using results from \TAC{}. Only such results are reported due to the page limit. 
We divide all imputers into three categories: (1) \CD{}, \LI{}, and \SL{} are \emph{traditional imputers} used in fingerprinting based indoor positioning~\cite{kaiser2014dealing,li2017turf,sorour2014joint}; (2) \MICE{} and \MF{} are \emph{autocorrelation based imputers} that exploit the autocorrelation of radio map records; and (3) \BRITS{}, \SSGAN{}, and \BISIMFamily{} are \emph{neural network imputers} that learn and utilize sequential data dependencies.

\ptitle{Implementation}
For \BRITS{}, \SSGAN{} and \BISIMFamily{}, we set the learning rate to $0.001$, the batch size to $32$, and the training epochs to $500$. The latent vector lengths in encoder/decoder are set to $64$. The length $T$ of an input feature sequence is tuned optimally to $5$. Longer sequences are sliced before encoding and assembled after decoding.
The Adam optimizer is used; all neural networks are tuned to optimal for evaluations.

\ptitle{Parameters}
We study how data imputers are affected by the sparsity of radio map. We introduce a
\emph{removal ratio} $\beta \in \{0, 10, 20, 30, 40, 50\}$\%---the fraction $\beta$ of RSSIs (or RPs) are randomly removed in the original radio map. The removed values serve as the ground-truth for measuring the imputation errors (metrics to be given below).
Here, $\beta$ carries a different meaning from the one (i.e., $\alpha$) used in Section~\ref{ssec:evaluation_differentiation}: the removal in this section is conducted after filling in all \MNAR{}s with $-100$ dBm.
In addition, we scale the original RP density from 60\% to 100\% such that we only keep $\{60, \ldots, 100\}$\% of RPs in the raw walking survey record table.

\ptitle{Metrics}
In addition to \APE, we consider the errors of the imputed results with respect to their ground-truth.
Specifically, we use the \emph{Mean Absolute Error} (MAE) for the $D$-dimensional fingerprints and the \emph{Euclidean Distance} for the 2-dimensional RPs.
In the subsequent reporting, we highlight the \best{best} and \secBest{second-best} imputation errors in each group of experiments.

\subsubsection{Results}

\ptitle{Accuracy Comparison}
We employ three location estimation algorithms: \KNN{}~\cite{zeinalipour2017anatomy}, \WKNN{}~\cite{fang2008location}, and random forest (\RF{})~\cite{jedari2015wi}.
Referring to Table~\ref{tab:pos_algs}, on both venues, \BISIMFamily{} imputers always clearly outperform the competitors across different location estimation algorithms.
This shows that the \BISIM{} data imputer contributes greatly to improving the indoor positioning accuracy.

In addition, \TBISIM{} performs better than \ABISIM{}, which shows the superiority of \TAC{}.
Both \BRITS{} and \SSGAN{} perform poorer than \BISIMFamily{} as they fail to capture the dependencies between fingerprints and RPs, which are handled by the encoder-decoder in \BISIMFamily{}.
Overall, neural network imputers perform much better than traditional imputers and autocorrelation based imputers.
The poor performance of autocorrelation based imputers is attributed to their inability to deal with heterogeneous radio map records.
%

Comparing the three location estimation algorithms, \WKNN{} performs best in most cases. 
%
In subsequent experiments, we thus use \WKNN{} for location estimation.

\begin{table*}
\footnotesize
\centering
\caption{Overall \APE{} Comparison (unit: meter)}\label{tab:pos_algs}
\setlength\tabcolsep{2.4pt}
\begin{tabular}{c|ccccccccc|ccccccccc}
\toprule
\multirow{2}{*}[-0.5em]{\begin{tabular}{@{}c@{}}location\\estimation alg.\end{tabular}}
& \multicolumn{9}{c|}{\SIV{}}                         & \multicolumn{9}{c}{\SV{}}                \\ 
\cmidrule{2-19}
 & \CD{}    & \LI{}   & \SL{}   & \MICE{}  & \MF{}    & \BRITS{} & \SSGAN{}  & \ABISIM{}    & \TBISIM{}        & \CD{}    & \LI{}   & \SL{}   & \MICE{}  & \MF{}    & \BRITS{} & \SSGAN{} & \ABISIM{}    & \TBISIM{}        \\ \midrule
$K$NN         & 6.79  & 5.76 & 6.83 & 15.37 & 15.58 & 2.99  & 2.26 & \secBest{1.98} & \best{1.78}  & 12.73 & 9.96 & 8.63 & 25.13 & 28.23 & 5.14 &  4.62 & \secBest{3.41} & \best{2.43}    \\ 
W$K$NN      & 6.64  & 5.76 & 7.10  & 15.37 & 15.65 & 3.07  & 2.23
 & \secBest{1.96} & \best{1.66} & 12.52 & 9.95 & 8.45 & 27.91 & 28.35 & 4.78 & 3.47 & \secBest{3.27} & \best{2.41}    \\ 
RF          & 7.23  & 5.57 & 7.35 & 15.00  & 15.36 & 5.07  & 4.49
  &\secBest{2.93} & \best{2.70}   & 11.28 & 9.25 & 9.03 & 26.81 & 27.64 & 18.52 & 8.02 & \secBest{3.44} & \best{3.10}   \\ \bottomrule
\end{tabular}
\vspace*{-10pt}
\end{table*}

\ptitle{Imputation Time Cost Comparison} 
The total time costs to impute the radio map are given in Table~\ref{tab:time_cost}. 
Traditional imputers, \LI{} and \SL{}, take much less time due to their simplicity.
\MICE{} and \MF{} involve iterative processes on matrices and thus take more time. \MF{} is the most time-consuming imputer as the high data sparsity of matrices makes it hard for \MF{}  to converge.
Next, \BRITS{} and \BISIMFamily{} take  time cost comparable to \MICE{} and \MF{}, but achieve much higher accuracy than all other models (cf.\ Table~\ref{tab:pos_algs}).
\SSGAN{} is the slowest among neural network-based imputers as its GAN model converges slowly~\cite{mescheder2018training}.
The most accurate imputer, \TBISIM{}, takes two minutes more than \BRITS{}  in imputation, while achieving an \APE{} improvement of $1$ m on both venues.
Considering that imputation is an offline procedure, employing \TBISIM{} is the most cost-effective.


\begin{table}[]
\footnotesize
    \centering
    {\setlength\tabcolsep{2.4pt} 
       \caption{Data Imputation Time Cost (unit: minute)}
    \label{tab:time_cost}
    \begin{tabular}{c|cccccccc}
    \toprule 
    & \LI{} & \SL{} & \MICE{} & \MF{} & \BRITS{} & \SSGAN{} & \ABISIM{} & \TBISIM{}   \\
    \midrule 
         \SIV{} &  1.35&2.41&12.06 &29.89 &13.84&21.41 &12.87&15.10  \\
        \midrule
         \SV{} &  2.38&5.50 &22.64 &67.02&23.13&33.43 & 22.74&25.43    \\
    \bottomrule
    \end{tabular}
 }
\end{table}

\begin{table}[!htbp]
\footnotesize
\centering
{\setlength\tabcolsep{1.0pt} 
\caption{APE on Bluetooth Data (unit: meter)}
    \label{tab:bt_ape}
\begin{tabular}{c|ccccccccc} 
\toprule
     & CD    & LI    & SL    & MICE  & MF    & BRITS & SSGAN & D-BiSIM & T-BiSIM \\
    \midrule
KNN  & 22.65 & 17.99 & 20.42 & 57.41 & 19.57 & 7.52&6.67  & \secBest{6.28}    & \best{5.95}    \\
WKNN & 22.76 & 16.14 & 18.7  & 57.27 & 19.68 & 7.33&6.74  & \secBest{6.24}    & \best{5.86}    \\
RF   & 23.21 & 17.69 & 20.7  & 63.37 & 20.36 & 9.49&8.31  & \secBest{7.13}    & \best{6.29}   \\
\bottomrule
\end{tabular}}
\end{table}

\begin{figure}[t]
\centering
\includegraphics[width=\columnwidth]{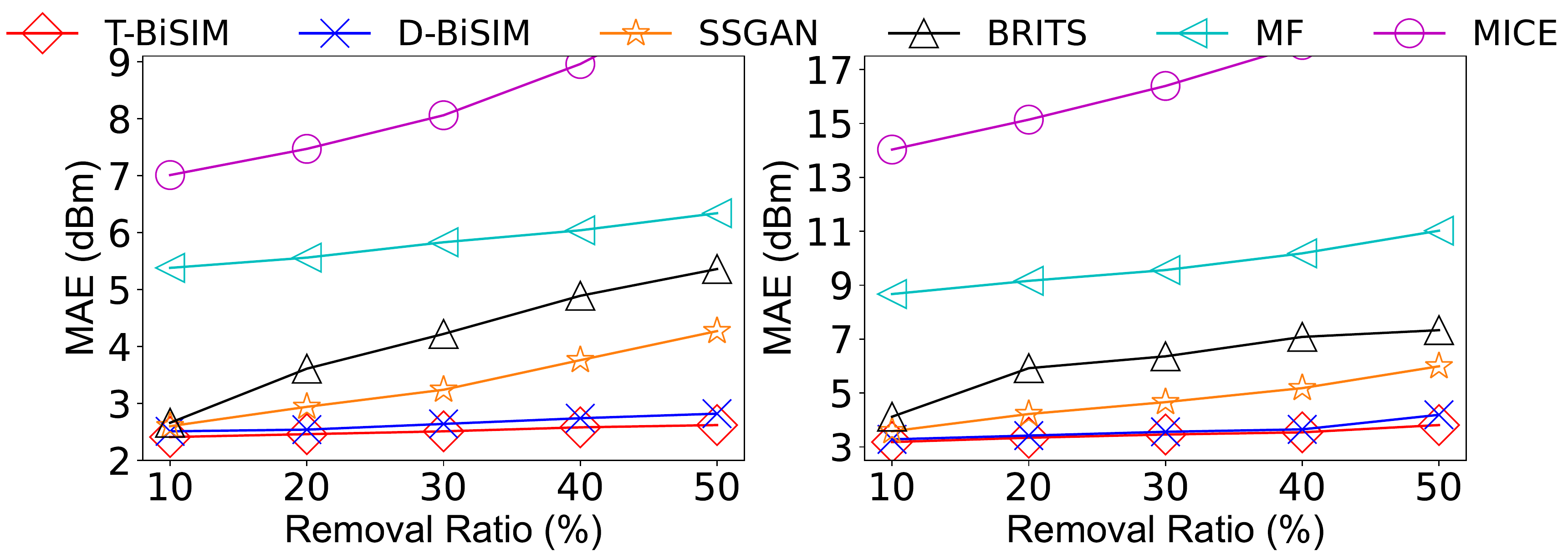}\\
{\small \ \ \ \ (a) \SIV{} \ \ \ \  \ \ \ \ \ \ \  \ \ \ \ \ \ \ \ \ \ \ \ \ \ \ \ \ \ \ (b) \SV{} }
\caption{The removal ratio $\beta$ vs. MAE.}
\label{fig:mae_vary_beta}
\vspace*{-10pt}
\end{figure}

\ptitle{Effect of Removal Ratio $\beta$}
We consider the imputation of RSSIs and RPs, respectively. 
Referring to Fig.~\ref{fig:mae_vary_beta}, when more RSSIs are removed from the radio map (due to a higher removal ratio), each method's MAE increases, as more missing values have to be imputed.
Still, \TBISIM{} and \ABISIM{} perform the best and second best in all tests, respectively, and their performance is affected the least by an increasing $\beta$.
The MAE of \MICE{} and \MF{} increase rapidly as their captured autocorrelation becomes less reliable when more RSSIs are removed.  
We disregard all traditional imputers from the  RSSI imputation comparison as they fill in $-100$ dBm by default.


\begin{figure}[t]
\centering
\includegraphics[width=\columnwidth]{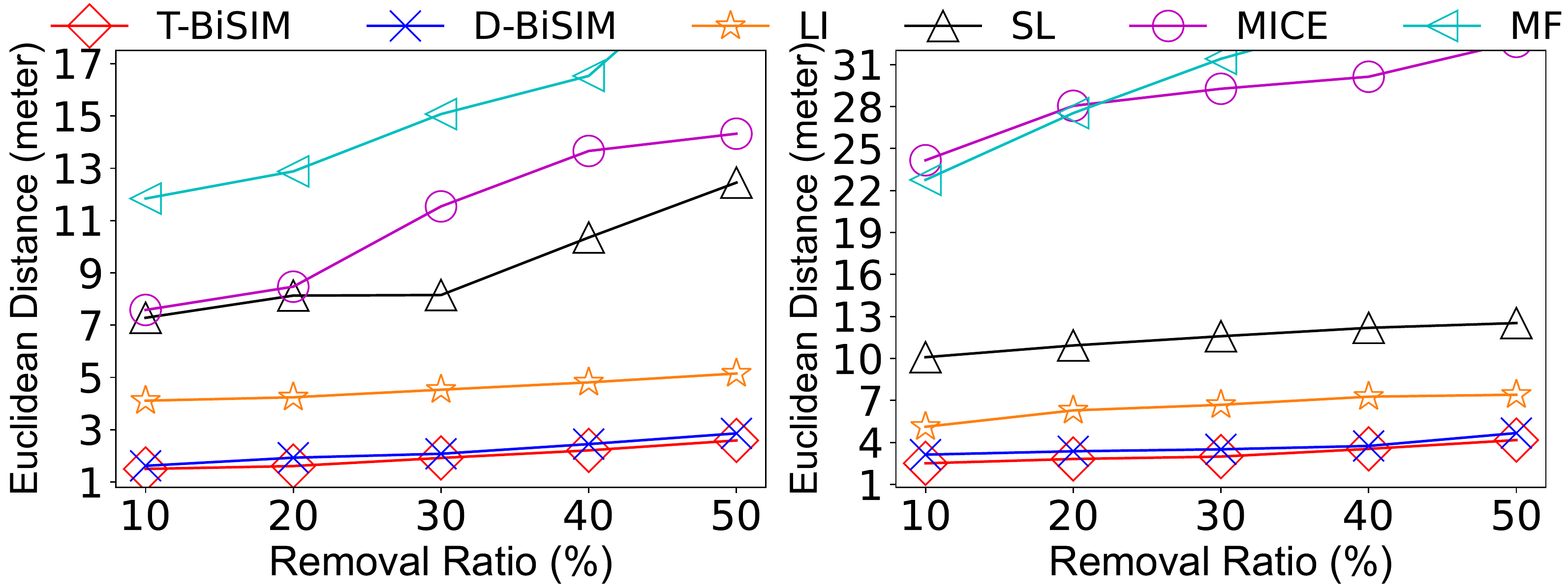}\\
\vspace*{-4pt}
{\small\ \  (a) \SIV{} \ \ \ \  \ \ \ \ \ \ \ \ \ \ \ \ \ \ \ \ \ \ \ \ \ \ \ (b) \SV{} }
\caption{The removal ratio $\beta$ vs. Euclidean distance.}
\label{fig:eu_vary_beta}
\vspace*{-10pt}
\end{figure}

Referring to Fig.~\ref{fig:eu_vary_beta}, for all imputers, the Euclidean distance error on RPs increases when more RPs are removed before imputation.
Still, \BISIMFamily{} is the best. When $50$\% of RPs are removed, \TBISIM{} retains a distance of $2.59$ ($4.16$) meters in \SIV{} (\SV{}), so it is robust to RP data sparsity.
We omit \CD{}, \BRITS{}, and \SSGAN{} for not involving RP imputation.

\begin{figure}
\centering
\begin{minipage}[t]{0.231\textwidth}
\centering
\includegraphics[width=\textwidth]{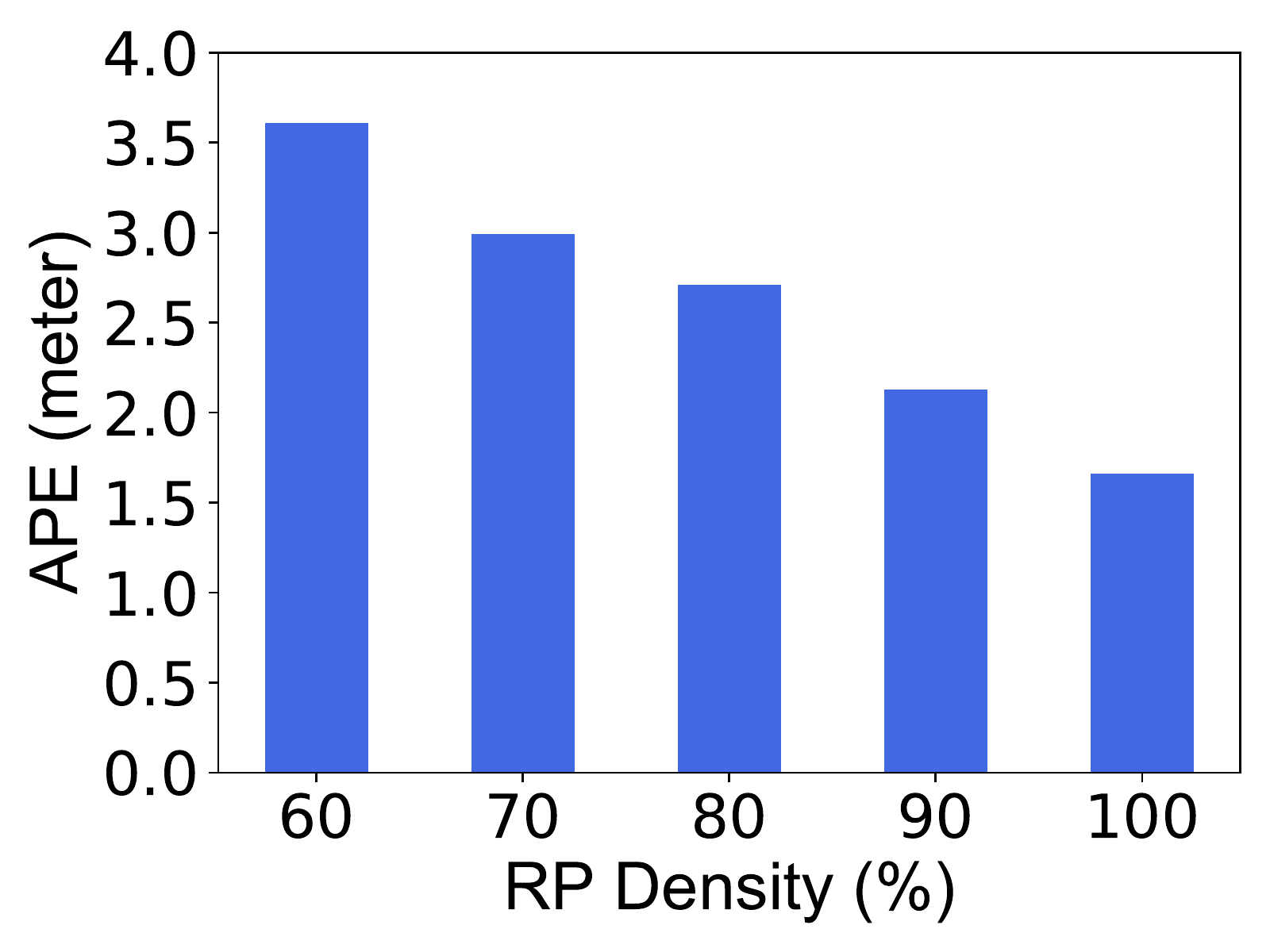}
\end{minipage}
\begin{minipage}[t]{0.231\textwidth}
\centering
\includegraphics[width=\textwidth]{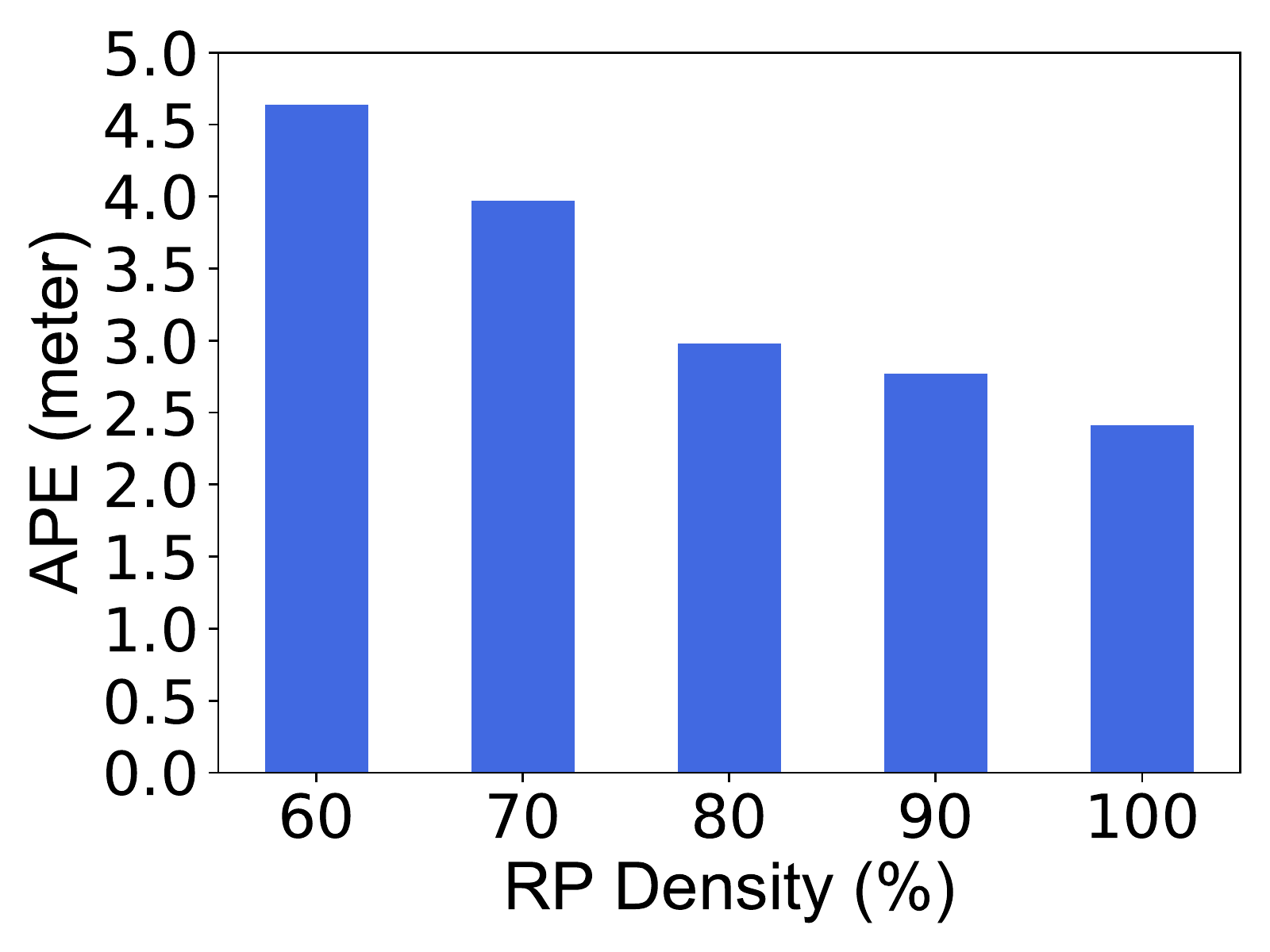}
\end{minipage}\\
{\small\ \ \ \ \  (a) \SIV{} \ \ \ \  \ \ \ \ \ \ \ \ \  \ \ \ \ \ \ \ \ \ \ \ \ \ (b) \SV{} }
\ExpCaption{The RP density vs. \APE{}.}\label{fig:rp_density_effect}
\end{figure}

\ptitle{Effect of RP Density}
Referring to Fig.~\ref{fig:rp_density_effect}, as fewer RPs are removed during walking surveying, \APE{} improves for \TBISIM{} as more RP information is available for the differentiator and the imputer.
In particular, the differentiator \TAC{} benefits from more RPs that results in better clustering of AP profiles, while the imputer \BISIM{} captures the temporal dependencies better if radio map records are denser.
Also, we observe that \SIV{} constantly achieves better \APE{} than \SV{}. We believe this is because \SIV{} features denser RPs.

\begin{figure}
\centering
\begin{minipage}[t]{0.231\textwidth}
\centering
\includegraphics[width=\textwidth]{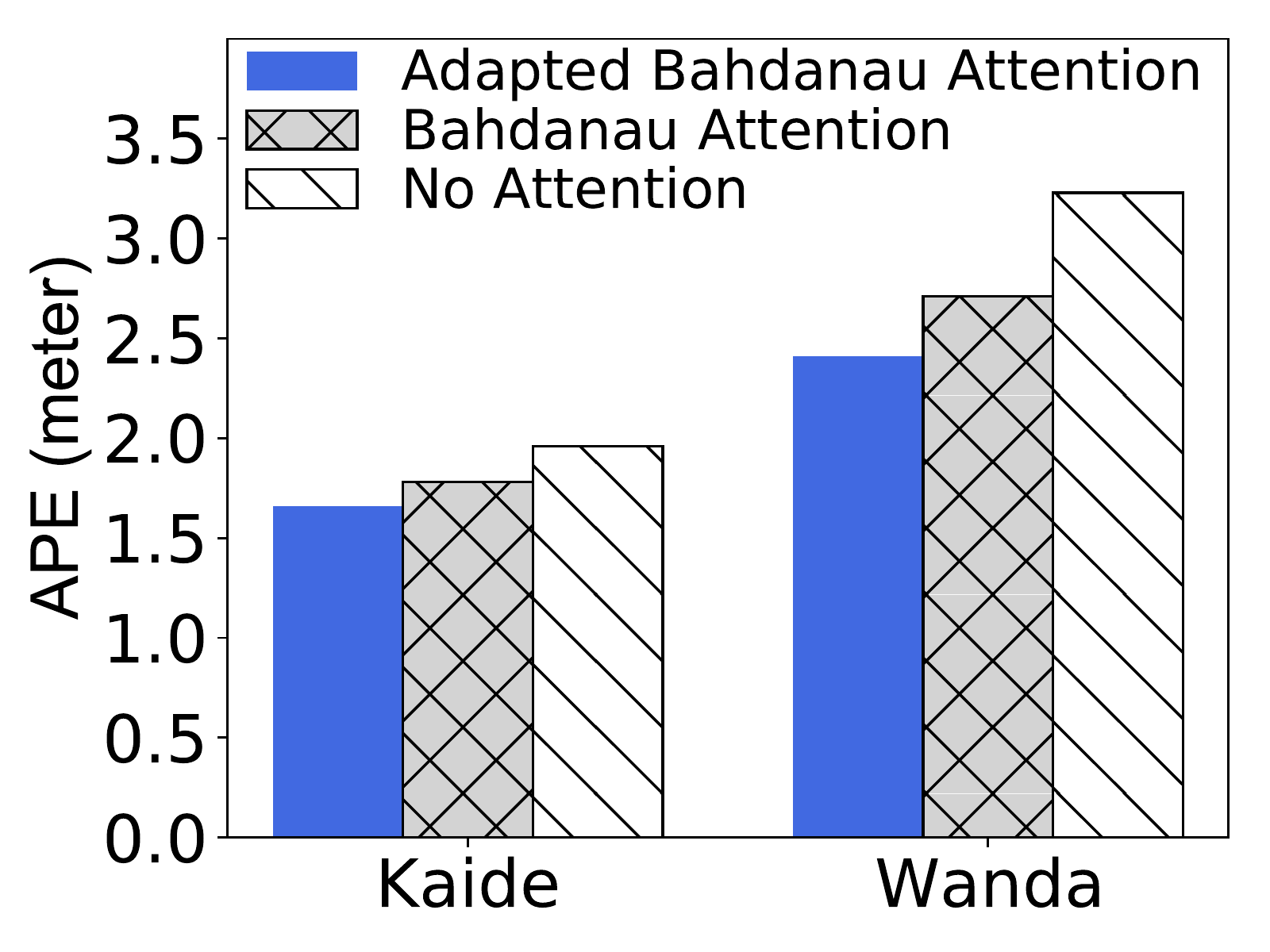}
\ExpCaption{Attention vs. \APE{}.}\label{fig:ablation_attention}
\end{minipage}
\begin{minipage}[t]{0.231\textwidth}
\centering
\includegraphics[width=\textwidth]{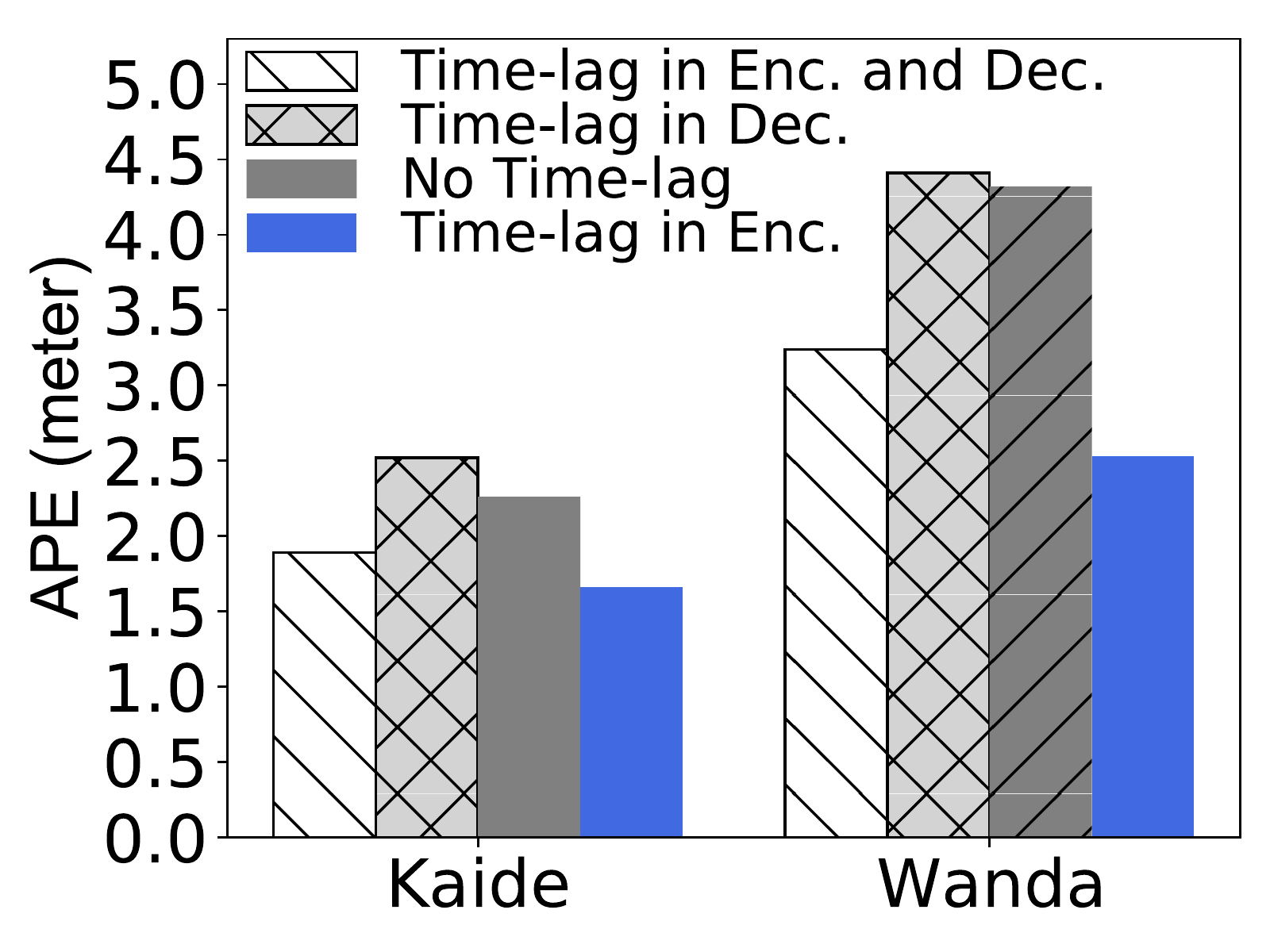}
\ExpCaption{Time-lag vs. \APE{}.}\label{fig:ablation_temporal_decay}
\end{minipage}
\end{figure}

\ptitle{Ablation Study (Attention)}
We compare the \TBISIM{} variants with (1) our adapted Bahdanau attention (Section~\ref{ssec:bisim_internals}), (2) traditional Bahdanau attention, and (3) no attention.
Referring to Fig.~\ref{fig:ablation_attention}, the variant without attention performs worst, showing the effectiveness of adding an attention unit in the encoder-decoder architecture.
Moreover, our adapted Bahdanau attention outperforms the traditional Bahdanau attention on both venues. This is because the adapted attention design focuses on the observed part of the input features and generates more accurate weights for imputation.

\ptitle{Ablation Study (Time-lag)}
Recall that Section~\ref{ssec:bisim_input} introduces a time-lag mechanism into \BISIM{}.
We compare the \TBISIM{} variants with (1) time-lag employed in encoders (fingerprint part) only (our design), (2) time-lag employed in decoders (RP part) only, (3) time-lag employed in both encoders and decoders, and (4) no time-lag employed.
Fig.~\ref{fig:ablation_temporal_decay} shows that our design with time-lag fingerprint vectors performs best and the variant without time-lag yields the highest \APE{}.
Interestingly, using time-lag vectors in both encoders and decoders degrades the performance. The possible reason is that the extra time-lag mechanism applied to decoders complicates the model and reduces its generalizability.

\ptitle{Generalizability}
We conduct additional experiments with Bluetooth fingerprinting data in a third venue (i.e., Longhu) to study the generalizability of our proposals. The APE results for the Bluetooth dataset from  \SVI{} are presented in Table~\ref{tab:bt_ape}. We see that \BISIMFamily{} continues to outperform the other data imputation methods with a significant advantage, indicating that the proposed imputation framework is effective in Bluetooth fingerprinting scenarios~\cite{iglesias2012indoor} and has the potential for applications across diverse indoor positioning systems.

\section{Related Work}
\label{sec:related}


\begin{table}[t]
    \centering
    \footnotesize
    {\setlength\tabcolsep{4pt} %
    \caption{Neural Networks for Time-series Imputation}\label{tab:MTSI}
    \begin{tabular}{@{}c|c|c|c|c|c@{}}
    \toprule
        Models & Features &  Labels   & Structure    &Time-lag  & Attention \\
    \midrule
     \BISIM{} & Imputed  &Imputed    & Seq2Seq & \checkmark & \checkmark  \\
    GRUD~\cite{che2018recurrent}& Imputed &-& RNN   & \checkmark & -  \\
    
    MRNN~\cite{yoon2017multi}&Imputed &-   & RNN   & \checkmark & -  \\
    
         BRITS~\cite{cao2018brits}&Imputed  & -   & RNN   & \checkmark & -  \\
         
         DeepMVI~\cite{bansal2021missing}&Imputed  & -   & Transformer   & - & \checkmark  \\
          GRIL~\cite{cini2021multivariate}&Imputed  &-    & GRU-GNN    & - & -  \\
          \midrule
          GRU-GAN~\cite{luo2018multivariate}&Imputed &-    & GRU-GAN    & \checkmark & -  \\
          NAOMI~\cite{liu2019naomi}& Imputed  &-   & GRU-GAN    & - & -  \\
          E2GAN~\cite{luo2019e2gan}& Imputed  &-   & GRU-GAN    & \checkmark & -  \\
           SSGAN~\cite{miao2021generative}& Imputed  & -   & RNN-GAN    & \checkmark & -  \\
           
    \bottomrule
    \end{tabular}}
\end{table}

\ptitle{Radio Map Completion}
Traditional positioning methods~\cite{firdaus2019accurate,dong2017dealing,gorak2018automatic} simply replace \NULL{} RSSIs in fingerprints with the minimum value of $-100$ dBm.
However, this adds errors to a radio map as missing RSSIs may be caused by random events (e.g., temporarily blocked signal transmission) and their actual values are not \NULL{} or $-100$ dBm.
Next, existing studies handle missing RPs based on a simple deletion of corresponding pairs~\cite{kaiser2014dealing}, linear interpolation with contextual RPs~\cite{li2017turf}, or semi-supervised learning using records with observed RPs~\cite{sorour2014joint}.
A major issue of linear interpolation and semi-supervised learning is that a radio map itself is sparse.
%
Differently, our solution differentiates \MAR{}s and \MNAR{}s and employs a sequential neural network to impute missing RSSIs and RPs jointly based on temporal dependencies of radio map records and correlations between fingerprints and RPs.

\ptitle{Missing Data Imputation}
Straightforward zero and mean filling approaches  usually yield low accuracy.
Autocorrelation-based imputation methods such as MICE~\cite{azur2011multiple} and MF~\cite{hastie2009elements, khayati2014memory} focus on homogeneous data records and do not fit in our setting where each record consists of a signal vector and a location. Moreover, these methods do not contend well with high data sparsity.
In addition, there exist deep learning studies for imputating multi-variate time series data~\cite{che2018recurrent, yoon2017multi,cao2018brits, miao2021generative, cini2021multivariate,bansal2021missing,khayati2020mind}. For instance,
Che et al.~\cite{che2018recurrent} incorporate masking and time-lag mechanisms into a vanilla GRU and impute \NULL{}s based on a weighted combination of the last observation and a global mean.
Relaxing smooth assumptions~\cite{che2018recurrent}, Cao et al.~\cite{cao2018brits} propose BRITS, a bidirectional RNN that regards missing values as trainable variables and imputes them directly by backpropagation of loss computed on observed data.
Moreover, generative adversarial networks (GANs)~\cite{luo2018multivariate, luo2019e2gan,liu2019naomi,miao2021generative} have been used to learn the overall distribution of a time-series dataset to impute missing values.
Table~\ref{tab:MTSI} compares these works.

Existing neural network approaches do not apply  to our problem setting directly.
First, existing models impute missing values in feature sequences only, while our problem needs to handle {missing} values in both feature sequences (missing RSSIs) and label sequences (missing RPs).
Second, existing models either disregard labels~\cite{che2018recurrent,yoon2017multi,bansal2021missing,cini2021multivariate} or assume a many-to-one setting where a time series corresponds to a single label~\cite{cao2018brits,miao2021generative,bansal2021missing}, while our problem is a many-to-many setting, where each fingerprint is associated with one RP.
Third, GAN-based models assume that all \NULL{}s are \MAR{}s, while our missing RSSIs form a mix of MARs and MNARs.
All these key differences call for a way to differentiate types of missing RSSIs and means of handling {missing} values in both features and labels jointly.

Differentiating \MAR{}s and \MNAR{}s is beneficial to missing data imputation.
Studies~\cite{little2019statistical,rubin1976inference,sterner2011missing} point out that the design of differentiation methods requires domain  knowledge, as data characteristics are  tied closely to the specific application.
Some studies focus on  differentiation methods for specific domains, e.g., longitudinal clinical trials~\cite{jin2022hybrid} and answer quality in surveys~\cite{chen2011finding}, but such studies are inapplicable in our data setting.
To the best of our  knowledge, we are the first to study the differentiation of missing RSSI values.

\ptitle{Indoor Positioning Data Cleansing}
Some studies~\cite{jeffery2006adaptive,chen2010leveraging,fazzinga2016exploiting,baba2013spatiotemporal,baba2016learning} use sensor deployment knowledge and time-series dependencies to repair missing readings caused by sensor failures. 
 Missing values in these studies are identifiers of sensors such as  RFID readers. 
In addition, 
Lin et al.~\cite{lin2020locater} propose a semi-supervised scheme to detect and impute missing AP identifiers in raw Wi-Fi connectivity data. Our work differs from these works in that we aim to impute numerical values instead of AP or RFID reader identifiers. Also, our radio map imputation targets fingerprinting-based localization at a point level rather than at a regional level.
Sun et al.~\cite{sun2021data} propose a sequential alignment-and-matching method to complete missing RSSI values and an AP distribution-based mapping method to amend missing and false location labels.
That work assumes that all \NULL{}s are \MNAR{}s and location labels are at the room level.
Therefore, it is not applicable to our problem.

    
    
         

    

\section{Conclusion}
\label{sec:conclusion}
We impute missing 
\textit{received signal strength indicator values} (RSSIs) and \textit{reference points} (RPs) in radio maps by designing a framework encompassing a missing RSSI differentiator and a data imputer. 
%
%
%
The clustering-based differentiator determines
missing at random (\MAR{}) and missing not at random (\MNAR{}),
whereas the model-based imputer leverages temporal dependencies and correlations in data to impute 
\MAR{}s 
and missing RPs. 
%
%
Extensive experimental studies demonstrate that our proposed framework clearly outperforms existing alternatives in terms of 
positioning and imputation accuracy.

In future work, 
it is of interest to design more efficient methods that enable online imputation of fingerprints. 
Also, it is relevant to integrate our separate differentiator and imputer into a single model, thus enabling end-to-end support of imputation processes.

\section*{Acknowledgements}\label{sec:ack}
This work is an extended version of the paper entitled "Data Imputation for Sparse Radio Maps in Indoor Positioning" published at ICDE 2023. The work was funded by Independent Research Fund Denmark
(No. 8022-00366B). Huan Li's work was supported by Aalborg University and EU MSCA programme (No. 882232). The work also benefited from discussions in the context of DIREC, a centre funded by the Innovation Fund Denmark.

\clearpage
\balance
\bibliographystyle{plain}
\bibliography{ref.bib}

\begin{thebibliography}{10}

\bibitem{indoorPositioning}
\url{https://www.researchandmarkets.com/reports/4765038/indoor-positioning-and-navigation-global-market}.

\bibitem{dBm}
\url{https://en.wikipedia.org/wiki/DBm}.

\bibitem{codeRepo}
\url{https://github.com/XLI-2020/BiSIM}.

\bibitem{dataset}
\url{https://www.kaggle.com/c/indoor-location-navigation}.

\bibitem{altman1994diagnostic}
Douglas~G Altman and J~Martin Bland.
\newblock Diagnostic tests. 1: Sensitivity and specificity.
\newblock {\em BMJ}, 308(6943):1552, 1994.

\bibitem{azur2011multiple}
Melissa~J Azur, Elizabeth~A Stuart, Constantine Frangakis, and Philip~J Leaf.
\newblock Multiple imputation by chained equations: What is it and how does it
  work?
\newblock {\em Int J Methods Psychiatr Res}, 20(1):40--49, 2011.

\bibitem{baba2016learning}
Asif~Iqbal Baba, Manfred Jaeger, Hua Lu, Torben~Bach Pedersen, Wei-Shinn Ku,
  and Xike Xie.
\newblock Learning-based cleansing for indoor {RFID} data.
\newblock In {\em SIGMOD}, pages 925--936, 2016.

\bibitem{baba2013spatiotemporal}
Asif~Iqbal Baba, Hua Lu, Xike Xie, and Torben~Bach Pedersen.
\newblock Spatiotemporal data cleansing for indoor {RFID} tracking data.
\newblock In {\em MDM}, pages 187--196, 2013.

\bibitem{bahdanau2014neural}
Dzmitry Bahdanau, Kyunghyun Cho, and Yoshua Bengio.
\newblock Neural machine translation by jointly learning to align and
  translate.
\newblock In {\em ICLR}, 2015.

\bibitem{bansal2021missing}
Parikshit Bansal, Prathamesh Deshpande, and Sunita Sarawagi.
\newblock Missing value imputation on multidimensional time series.
\newblock {\em Proc. VLDB Endow.}, 14(11):2533--2545, 2021.

\bibitem{cao2018brits}
Wei Cao, Dong Wang, Jian Li, Hao Zhou, Yitan Li, and Lei Li.
\newblock {BRITS}: Bidirectional recurrent imputation for time series.
\newblock In {\em NeurIPS}, pages 6776--6786, 2018.

\bibitem{chang2014crowdsourcing}
Kyungmin Chang and Dongsoo Han.
\newblock Crowdsourcing-based radio map update automation for {Wi-Fi}
  positioning systems.
\newblock In {\em Geocrowd}, pages 24--31, 2014.

\bibitem{che2018recurrent}
Zhengping Che, Sanjay Purushotham, Kyunghyun Cho, David Sontag, and Yan Liu.
\newblock Recurrent neural networks for multivariate time series with missing
  values.
\newblock {\em Sci. Rep.}, 8(1):1--12, 2018.

\bibitem{chen2010leveraging}
Haiquan Chen, Wei-Shinn Ku, Haixun Wang, and Min-Te Sun.
\newblock Leveraging spatio-temporal redundancy for {RFID} data cleansing.
\newblock In {\em SIGMOD}, pages 51--62, 2010.

\bibitem{chen2011finding}
Pu-Shih~Daniel Chen.
\newblock Finding quality responses: The problem of low-quality survey
  responses and its impact on accountability measures.
\newblock {\em Research in Higher Education}, 52(7):659--674, 2011.

\bibitem{cho2014properties}
Kyunghyun Cho, Bart van Merri{\"e}nboer, Dzmitry Bahdanau, and Yoshua Bengio.
\newblock On the properties of neural machine translation: Encoder--decoder
  approaches.
\newblock In {\em SSST-8}, pages 103--111, 2014.

\bibitem{cini2021multivariate}
Andrea Cini, Ivan Marisca, and Cesare Alippi.
\newblock Multivariate time series imputation by graph neural networks.
\newblock {\em arXiv preprint arXiv:2108.00298}, 2021.

\bibitem{dong2017dealing}
Kai Dong, Zhen Ling, Xiangyu Xia, Haibo Ye, Wenjia Wu, and Ming Yang.
\newblock Dealing with insufficient location fingerprints in {Wi-Fi} based
  indoor location fingerprinting.
\newblock {\em Wirel. Commun. Mob. Comput.}, 2017.

\bibitem{fang2008location}
Shih-Hau Fang, Tsung-Nan Lin, and Po-Chiang Lin.
\newblock Location fingerprinting in a decorrelated space.
\newblock {\em IEEE Trans Knowl Data Eng.}, 20(5):685--691, 2008.

\bibitem{fazzinga2016exploiting}
Bettina Fazzinga, Sergio Flesca, Filippo Furfaro, and Francesco Parisi.
\newblock Exploiting integrity constraints for cleaning trajectories of
  {RFID}-monitored objects.
\newblock {\em ACM Trans. Database Syst.}, 41(4):1--52, 2016.

\bibitem{firdaus2019accurate}
Firdaus Firdaus, Noor~Azurati Ahmad, and Shamsul Sahibuddin.
\newblock Accurate indoor-positioning model based on people effect and
  ray-tracing propagation.
\newblock {\em Sensors}, 19(24):5546, 2019.

\bibitem{garcia2010theoretical}
Vicente Garc{\'\i}a, Ramon~A Mollineda, and J~Salvador S{\'a}nchez.
\newblock Theoretical analysis of a performance measure for imbalanced data.
\newblock In {\em ICPR}, pages 617--620, 2010.

\bibitem{gorak2018automatic}
Rafa{\l} G{\'o}rak and Marcin Luckner.
\newblock Automatic detection of missing access points in indoor positioning
  system.
\newblock {\em Sensors}, 18(11):3595, 2018.

\bibitem{han2014kailos}
Dongsoo Han, Sangjae Lee, and Sunghoon Kim.
\newblock {KAILOS}: {KAIST} indoor locating system.
\newblock In {\em IPIN}, pages 615--619, 2014.

\bibitem{hastie2009elements}
Trevor Hastie, Robert Tibshirani, Jerome~H Friedman, and Jerome~H Friedman.
\newblock {\em The elements of statistical learning: Data mining, inference,
  and prediction}, volume~2.
\newblock 2009.

\bibitem{he2015wi}
Suining He and S-H~Gary Chan.
\newblock {Wi-Fi} fingerprint-based indoor positioning: Recent advances and
  comparisons.
\newblock {\em IEEE Commun. Surv. Tutor.}, 18(1):466--490, 2015.

\bibitem{iglesias2012indoor}
H{\'e}ctor Jos{\'e}~P{\'e}rez Iglesias, Valent{\'\i}n Barral, and Carlos~J
  Escudero.
\newblock Indoor person localization system through rssi bluetooth
  fingerprinting.
\newblock In {\em 2012 19th International Conference on Systems, Signals and
  Image Processing (IWSSIP)}, pages 40--43. IEEE, 2012.

\bibitem{jedari2015wi}
Esrafil Jedari, Zheng Wu, Rashid Rashidzadeh, and Mehrdad Saif.
\newblock {Wi-Fi} based indoor location positioning employing random forest
  classifier.
\newblock In {\em IPIN}, pages 1--5, 2015.

\bibitem{jeffery2006adaptive}
Shawn~R Jeffery, Minos Garofalakis, and Michael~J Franklin.
\newblock Adaptive cleaning for {RFID} data streams.
\newblock {\em Proc. VLDB Endow.}, 6:163--174, 2006.

\bibitem{jin2022hybrid}
Man Jin.
\newblock A hybrid return to baseline imputation method to incorporate mar and
  mnar dropout missingness.
\newblock {\em Contemporary Clinical Trials}, 120:106859, 2022.

\bibitem{jung2016performance}
Suk~Hoon Jung, Byeong-Cheol Moon, and Dongsoo Han.
\newblock Performance evaluation of radio map construction methods for {Wi-Fi}
  positioning systems.
\newblock {\em IEEE Trans. Intell. Transp. Syst.}, 18(4):880--889, 2016.

\bibitem{kaiser2014dealing}
Ji{\v{r}}{\'\i} Kaiser.
\newblock Dealing with missing values in data.
\newblock {\em J. Syst. Integr.}, 5(1), 2014.

\bibitem{kargar2021predict}
Saeed Kargar, Heiner Litz, and Faisal Nawab.
\newblock Predict and write: Using k-means clustering to extend the lifetime of
  {NVM} storage.
\newblock In {\em ICDE}, pages 768--779, 2021.

\bibitem{khalajmehrabadi2017modern}
Ali Khalajmehrabadi, Nikolaos Gatsis, and David Akopian.
\newblock Modern {WLAN} fingerprinting indoor positioning methods and
  deployment challenges.
\newblock {\em IEEE Commun. Surv. Tutor.}, 19(3):1974--2002, 2017.

\bibitem{khayati2014memory}
Mourad Khayati, Michael B{\"o}hlen, and Johann Gamper.
\newblock Memory-efficient centroid decomposition for long time series.
\newblock In {\em ICDE}, pages 100--111, 2014.

\bibitem{khayati2020mind}
Mourad Khayati, Alberto Lerner, Zakhar Tymchenko, and Philippe
  Cudr{\'e}-Mauroux.
\newblock Mind the gap: An experimental evaluation of imputation of missing
  values techniques in time series.
\newblock In {\em Proc. VLDB Endow.}, volume~13, pages 768--782, 2020.

\bibitem{li2017turf}
Chenhe Li, Qiang Xu, Zhe Gong, and Rong Zheng.
\newblock {TuRF}: Fast data collection for fingerprint-based indoor
  localization.
\newblock In {\em IPIN}, pages 1--8, 2017.

\bibitem{lin2020locater}
Yiming Lin, Daokun Jiang, Roberto Yus, Georgios Bouloukakis, Andrew Chio,
  Sharad Mehrotra, and Nalini Venkatasubramanian.
\newblock {Locater}: Cleaning {WiFi} connectivity datasets for semantic
  localization.
\newblock {\em Proc. VLDB Endow.}, 14(3):329--341, 2020.

\bibitem{little2019statistical}
Roderick~JA Little and Donald~B Rubin.
\newblock {\em Statistical analysis with missing data}, volume 793.
\newblock John Wiley \& Sons, 2019.

\bibitem{liu2019naomi}
Yukai Liu, Rose Yu, Stephan Zheng, Eric Zhan, and Yisong Yue.
\newblock {NAOMI}: Non-autoregressive multiresolution sequence imputation.
\newblock {\em Adv Neural Inf Process Syst.}, 32:11238--11248, 2019.

\bibitem{luo2018multivariate}
Yonghong Luo, Xiangrui Cai, Ying Zhang, Jun Xu, et~al.
\newblock Multivariate time series imputation with generative adversarial
  networks.
\newblock {\em Adv Neural Inf Process Syst.}, 31:1596--1607, 2018.

\bibitem{luo2019e2gan}
Yonghong Luo, Ying Zhang, Xiangrui Cai, and Xiaojie Yuan.
\newblock {E2GAN}: End-to-end generative adversarial network for multivariate
  time series imputation.
\newblock In {\em IJCAI}, pages 3094--3100, 2019.

\bibitem{mescheder2018training}
Lars Mescheder, Andreas Geiger, and Sebastian Nowozin.
\newblock Which training methods for gans do actually converge?
\newblock In {\em International conference on machine learning}, pages
  3481--3490. PMLR, 2018.

\bibitem{miao2021generative}
Xiaoye Miao, Yangyang Wu, Jun Wang, Yunjun Gao, Xudong Mao, and Jianwei Yin.
\newblock Generative semi-supervised learning for multivariate time series
  imputation.
\newblock In {\em AAAI}, volume~35, pages 8983--8991, 2021.

\bibitem{pulkkinen2011semi}
Teemu Pulkkinen, Teemu Roos, and Petri Myllym{\"a}ki.
\newblock Semi-supervised learning for {WLAN} positioning.
\newblock In {\em ICANN}, pages 355--362, 2011.

\bibitem{quezada2022data}
Darwin Quezada-Gaibor, Lucie Klus, Joaqu{\'\i}n Torres-Sospedra, Elena~Simona
  Lohan, Jari Nurmi, Carlos Granell, and Joaqu{\'\i}n Huerta.
\newblock Data cleansing for indoor positioning {Wi-Fi} fingerprinting
  datasets.
\newblock In {\em MDM}, pages 367--371, 2022.

\bibitem{rubin1976inference}
Donald~B Rubin.
\newblock Inference and missing data.
\newblock {\em Biometrika}, 63(3):581--592, 1976.

\bibitem{sadowski2018rssi}
Sebastian Sadowski and Petros Spachos.
\newblock {RSSI}-based indoor localization with the {Internet of Things}.
\newblock {\em IEEE Access}, 6:30149--30161, 2018.

\bibitem{sorour2014joint}
Sameh Sorour, Yves Lostanlen, Shahrokh Valaee, and Khaqan Majeed.
\newblock Joint indoor localization and radio map construction with limited
  deployment load.
\newblock {\em IEEE Trans. Mobile Comput.}, 14(5):1031--1043, 2014.

\bibitem{sterner2011missing}
William~R Sterner.
\newblock What is missing in counseling research? reporting missing data.
\newblock {\em Journal of Counseling \& Development}, 89(1):56--62, 2011.

\bibitem{sun2020wifi}
Haotai Sun, Xiaodong Zhu, Yuanning Liu, and Wentao Liu.
\newblock Wifi based fingerprinting positioning based on seq2seq model.
\newblock {\em Sensors}, 20(13):3767, 2020.

\bibitem{sun2021data}
Jing Sun, Bin Wang, Xiaoxu Song, and Xiaochun Yang.
\newblock Data cleaning for indoor crowdsourced {RSSI} sequences.
\newblock In {\em APWeb-WAIM}, pages 267--275, 2021.

\bibitem{wang2017research}
Pengfei Wang and Yufeng Luo.
\newblock Research on wifi indoor location algorithm based on rssi ranging.
\newblock In {\em ICISCE}, pages 1694--1698, 2017.

\bibitem{wu2014smartphones}
Chenshu Wu, Zheng Yang, and Yunhao Liu.
\newblock Smartphones based crowdsourcing for indoor localization.
\newblock {\em IEEE Trans. Mobile Comput.}, 14(2):444--457, 2014.

\bibitem{xu2015enhancing}
Han Xu, Zheng Yang, Zimu Zhou, Longfei Shangguan, Ke~Yi, and Yunhao Liu.
\newblock Enhancing {WiFi}-based localization with visual clues.
\newblock In {\em UbiComp}, pages 963--974, 2015.

\bibitem{yoon2017multi}
Jinsung Yoon, William~R Zame, and Mihaela van~der Schaar.
\newblock Multi-directional recurrent neural networks: A novel method for
  estimating missing data.
\newblock In {\em ICML Time Series Workshop}, 2017.

\bibitem{zeinalipour2017anatomy}
Demetrios Zeinalipour-Yazti and Christos Laoudias.
\newblock The anatomy of the anyplace indoor navigation service.
\newblock {\em SIGSPATIAL Special}, 9(2):3--10, 2017.

\end{thebibliography}

\end{document}